\documentclass[a4paper,fleqn,usenatbib]{mnras}

\usepackage[T1]{fontenc}
\usepackage{ae,aecompl}

\usepackage{graphicx}    
\usepackage{amsmath}    
\usepackage{amssymb}    
\usepackage{subcaption}
\usepackage{booktabs}
\usepackage[flushleft]{threeparttable}
\usepackage{amsmath}
\usepackage{longtable}
\usepackage{multirow}

\newcommand{\mathsc}[1]{{\text{\normalfont\scshape#1}}} 
\newcommand\ho{\ifmmode {\rm H\hspace{.1em}\mathsc{i}} \else \mbox{\rm H\,\scshape{i}} \fi}
\newcommand{\kms}{\ifmmode {\rm km\ s}^{-1} \else km s$^{-1}$ \fi}

\newcommand{\msun}{M_{\odot}}
\newcommand{\hi}{\mbox{\rm H$\,$\scshape{i}}}

\title{Observations of cold gas and star formation in dwarf S0 galaxies}

\author[X. Ge et al.]{
Xue Ge$^{1,3}$,
Qiu-Sheng Gu\thanks{Corresponding author Email: qsgu@nju.edu.cn}$^{2,3}$,
Rub\'en Garc\'ia-Benito$^{4}$,
Shi-Ying Lu$^{2,3}$,
Cheng-Long Lei$^{1}$,
\newauthor
and Nan Ding\thanks{Corresponding author Email: orient.dn@smail.nju.edu.cn}$^{5}$
\\
$^{1}$School of Physics and Electronic Engineering, Jiangsu Second Normal University, Nanjing, Jiangsu 211200, China\\
$^{2}$School of Astronomy and Space Science, Nanjing University, Nanjing, Jiangsu 210093, China\\
$^{3}$Key Laboratory of Modern Astronomy and Astrophysics (Nanjing University), Ministry of Education, Nanjing 210093, China\\
$^{4}$Instituto de Astrof\'isica de Andaluc\'ia (CSIC), P.O. Box 3004, 18080 Granada, Spain\\
$^{5}$School of Physical Science and Technology, Kunming University, Kunming 650214, P. R. China
}

\date{Accepted xxx. Received xxx; in original form xxx}

\pubyear{2020}

\begin{document}
\label{firstpage}
\pagerange{\pageref{firstpage}--\pageref{lastpage}}
\maketitle

\begin{abstract}
Very little work has been done on star formation in dwarf lenticular galaxies (S0s).
We present  2D-spectroscopic and millimetre observations made by Centro Astron\'omico Hispano Alem\'an (CAHA) 3.5 m optical and the IRAM-30 m millimetre telescopes, respectively, for a sample of four dwarf S0 galaxies with multiple star formation regions in the field environment. We find that although most of the sources deviate from the star forming main sequence relation, they all follow the Kennicutt-Schmidt law. 
After comparing the stellar and H$\alpha$ kinematics, we find that the velocity fields of both stars and ionized gas do not show regular motion and the velocity dispersions of stars and ionized gas are low in the regions with high star formation, suggesting these star-forming S0 galaxies still have significant rotation. This view can be supported by the result that most of these dwarf S0 galaxies are classified as fast rotators.
The ratio of average atomic gas mass to stellar mass ($\sim 47\%$) is much greater than that of molecular gas mass to stellar mass ($\sim 1\%$).
In addition, the gas-phase metallicities in the star-forming regions are lower than that of the non-star-forming regions.
These results indicate that the extended star formation may originate from the combination of abundant atomic hydrogen, long dynamic time scale and low-density environment.

\end{abstract}


\section{Introduction}  
\label{sec1}

Lenticular galaxies (S0s) are introduced as a distinct morphological type in the Hubble tuning fork diagram \citep{1936rene.book.....H}. They are classified as early-type galaxies (ETGs), a class of galaxies that have evolved passively after a big burst of star formation, and considered as an intermediate population between spirals and ellipticals. One of the most popular theories, the hierarchical two-phase accumulation, is usually used to explain the star formation history of ETGs \citep{2010ApJ...725.2312O}. The first phase is the main mechanism for stellar growth via gas collapse at high redshift, while the second stage is the size growth by the accretion of satellite systems at low redshift. Although this scenario can successfully explain the formation of many elliptical galaxies, the formation of S0 galaxies is not yet well understood.

Previous studies have found that there are many processes that can be introduced to explain the formation of S0 galaxies. One of them is the removal of cold gas from spiral galaxies via either secular evolution or environmental effects \citep{1996Natur.379..613M, 1998ApJ...495..139M, 1998ApJ...502L.133B, 2009MNRAS.398..312G, 2011MNRAS.415.1783B}. This mechanism is responsible for the properties of low star formation and prominent bulges in S0 galaxies. On the other hand, the results from numerical simulations have shown that the merger mechanism can also produce S0-like remnants \citep{2014A&A...565A..31T, 2015A&A...573A..78Q}. 
\cite{2015ApJ...804...32G} suggested that the compact spheroidal objects at high redshift can evolve into S0 galaxies through cold gas accretion. This accretion process at high redshift can increase the number of S0 galaxies seen in the present Universe. A recent study shows that violent disk instability could also be an important forming mechanism of S0 galaxies \citep{2018ApJ...862L..12S}.

The conventional perception is that the star formation rate (SFR) in S0 galaxies is low, which may be explained by two scenarios. The first is that S0 galaxies have little or no gas content. As found by \cite{1992ApJ...387..484B}, most S0 galaxies have lower $M(HI)/L_{B}$ than that of spiral galaxies. The second explanation for the low SFR is the morphological quenching, which is usually attributed to large and red bulges in their centers \citep{2009ApJ...707..250M, 2018ApJ...854..111X}.

However, many studies have shown that atomic and even molecular gas content is present in the volume-limited samples of S0 and/or elliptical galaxies \citep{2003ApJ...584..260W, 2006ApJ...644..850S, 2010ApJ...725..100W}. S0 galaxies could be rather rich in neutral hydrogen based on the observations of the 21 cm line \citep{1991A&A...243...71V}. 
\cite{2009AJ....138..579K} found that ETGs with star formation reside in similar regions of spirals in the color versus stellar mass space, and the fraction of ETGs with star formation increase with decreasing stellar mass. The variation in fraction of ETGs with star formation indicates that the origins of star-formation activity in ETGs might be different in different stellar mass ranges. High-mass star-forming S0 galaxies are often triggered by mergers, while low-mass S0 galaxies form stars by gas accretion. 

The kinematics of ETGs are closely related to the gas content. \cite{2016MNRAS.455..214D} and \cite{2019MNRAS.486.1404D} presented the cold gas observations of massive ETGs. They found that the slow rotators or the galaxies with higher velocity dispersion hold less gas. They attributed it to the gas-rich merger events, but the stellar mass-loss and hot halo cooling also play an important role. Similar results have been fond by \cite{2011MNRAS.414..940Y} who explored the relationship between the star formation history and the gas mass in ETGs using $\rm ATLAS^{3D}$ survey. They concluded that the smaller the angular momentum of the ETGs, the less the gas content, suggesting that the formation processes for slow rotators can effectively destroy the molecular gas or prevent the gas from accretion. However, \cite{2012MNRAS.422.1835S} and \cite{2014MNRAS.444.3388S} found that a large diversity of \hi\ masses and morphologies not only within fast rotators but also within slow rotators and the neutral hydrogen can  provide fresh gas for the residual star formation in ETGs.
Recent study also found that the external accretion of gas may fuel the star formation or an Active Galactic Nucleus \citep{2021arXiv210303277R}.

\begin{figure*}
\centering
\includegraphics[angle=0,width=0.2\textwidth]{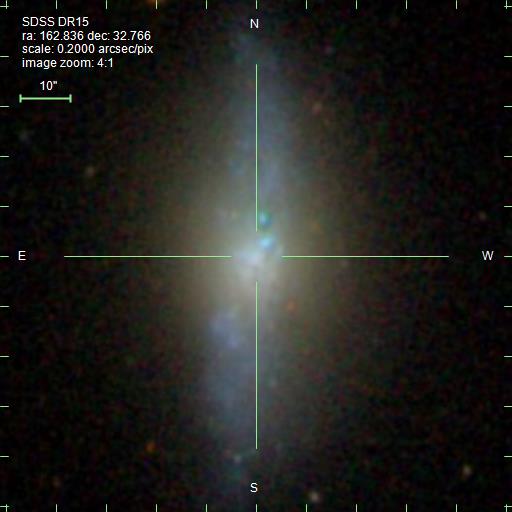}
\includegraphics[angle=0,width=0.2\textwidth]{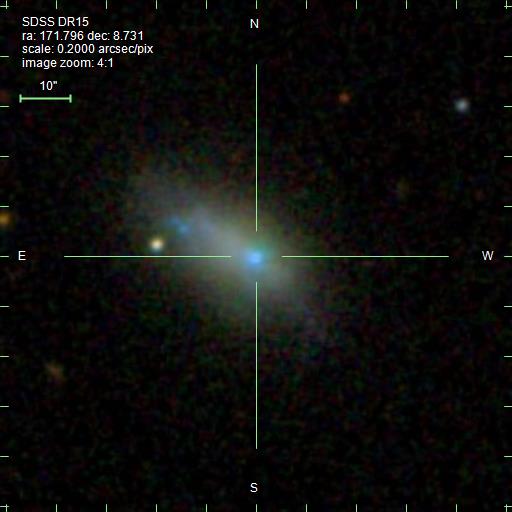}
\includegraphics[angle=0,width=0.2\textwidth]{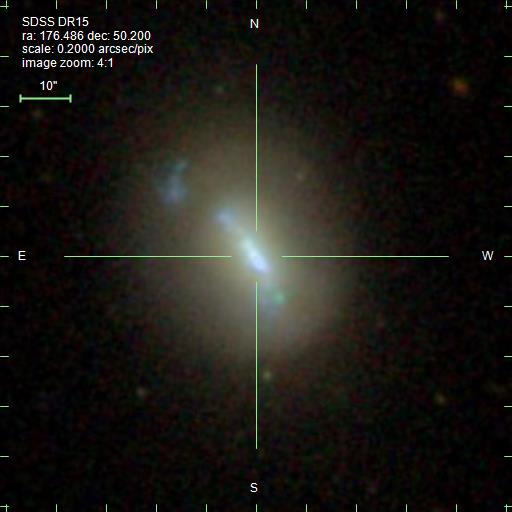}
\includegraphics[angle=0,width=0.2\textwidth]{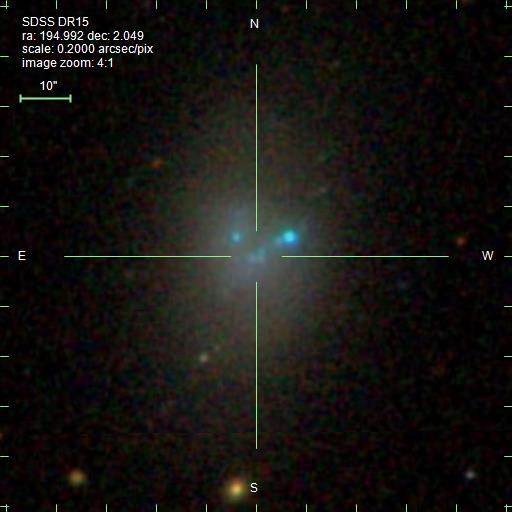}
\caption{SDSS composite color images for the four S0 galaxies. From left to right, they are PGC 32543, PGC 35225, PGC 36686 and PGC 44685, respectively.}
\label{f1}
\end{figure*}

\cite{2016ApJ...831...63X} studied the nuclear activities in nearby S0 galaxies using Sloan Digital Sky Survey Data Release 7 (SDSS DR7) and found that roughly 8 percent of S0 galaxies show central star-formation activity. It should be noted that the fraction might be higher considering that low level star formation which often present in objects with low-ionization nuclear emission-line regions may be missed if only through optical spectral analysis. Therefore, the number of star-forming S0 galaxies in \cite{2016ApJ...831...63X} should be considered a lower limit.
In order to understand the spatially resolved properties of star-forming S0 galaxies from \cite{2016ApJ...831...63X}, we have started a program to obtain Integral Field Spectroscopy (IFS) data with the Centro Astron\'omico Hispano Alem\'an (CAHA) 3.5 m telescope.  
Based on the star-forming S0 galaxies sample, we here present four low-mass cases with multiple nuclear structures.
Figure \ref{f1} shows their false-color (g-, r-, and i-bands) images from SDSS. We note that these S0 galaxies have stellar masses $M_{\ast}\lesssim 10^{9}M_{\odot}$.
In order to investigate the molecular gas content in the four S0 galaxies, we simultaneously observed the CO($J$=1-0) and CO($J$=2-1) with the IRAM 30-m telescope.
ETGs generally have large stellar masses. Many previous studies almost involved high-mass ($\rm \log\ M_{\ast}>10.0$) samples, and few studies focus on such low-mass star-forming S0 galaxies. In this work, we hope that the combination of IFS and millimeter of dwarf S0 galaxies can provide important clues to the star formation history of ETGs.

The paper is organized as follows. In Section \ref{sec2}, we show the observations and reductions of optical IFS and millimeter data. Section \ref{sec3} and \ref{sec4} present the results and discussion. Finally, we give the summary in Sections \ref{sec5}. In the paper, we assume constant cosmographic parameters where $\Omega_{M}$=0.3, $\Omega_{\Lambda}$=0.7 and $\rm H=70~km~s$$^{-1}$~Mpc$^{-1}$ and adopt a \cite{1955ApJ...121..161S} IMF.

\section{Observations and Data Reductions}
\label{sec2}

\begin{table*}
\setlength{\tabcolsep}{0.2mm}
\caption{A summary of star-forming S0 galaxies sample. Col (1) source name; Col (2-4) coordinates and redshift; Col (5) gas-phase metallicity; Col (6-7) stellar mass and SFR; Col (8-9) densities of SFR and gas; Col (10-11) luminosity of CO (J=1-0) and CO (J=2-1); Col (12-13) the masses of $\rm H_{2}$ and \hi.\\
a. The stellar mass is estimated by adopting pPXF.\\
b. All \hi\ masses, except for PGC 36686 \citep[adopted  from][]{2015MNRAS.447.1531C}, are adopted from \citet{2018ApJ...861...49H}.}
\centering 
\small
\label{tab1}
\begin{tabular}{cccccccccccccc}
\hline\hline
Source &   RA             & DEC        &   D & 12+$\rm \log(O/H)$ & $\rm log\ M_{\ast}^{a}$ & log\ SFR   &  $\log$ $\Sigma_{\rm SFR}$  & $\log$ $\Sigma_{\rm gas}$ &   $\log\,L^{\prime}_{\rm CO (J=1-0)}$ &    
$\log\,L^{\prime}_{\rm CO (J=2-1)}$ & $\log$ $\rm M_{H_{2}}$ & $\log \rm M_{\hi}^b$\\
            &  J2000.0       &  J2000.0        & Mpc  & & $ M_{\odot} $   &  $M_{\odot} \rm yr^{-1}$ &  $M_\odot\ \rm yr^{-1}\ kpc^{-2}$  &  $M_\odot\ \rm pc^{-2}$ &  K km s$^{-1}$ pc$^2$   &  K km s$^{-1}$ pc$^2$   &   $M_{\odot}$  &   $M_{\odot}$ \\
 (1) & (2)  & (3)  & (4)  & (5)  & (6)   & (7)   & (8)   & (9) & (10) & (11) & (12) & (13) \\
\hline
PGC 32543    &  162.836 & 32.766  & 6.36  & 8.54     &  8.62    & -1.39 & -1.54  & 1.21 & 5.88  & 5.75  & 6.51 & 8.45\\
PGC 35225    &  171.795 & 8.731  &  10.35  & 8.54      &  8.32    & -1.02 & -1.43  & 1.20 & 5.86  & 5.70    & 6.49 & 7.83\\
PGC 36686    &  176.485 & 50.199  &  7.44  & 8.41      &  8.86    & -1.48 &  -1.41 & 1.35 & 6.22  & 5.85   & 6.85 & 8.25\\
PGC 44685    &  194.990 & 2.050  &  8.76  & 8.37      &  8.46    & -1.19 &  -1.19  & 1.40 & 5.94  & 5.42  & 6.57 & 8.27\\
\hline\hline
\end{tabular}
\end{table*}

\begin{table*}
\small
\label{tab2}
\caption{The summary of fitting parameters. Col.(1): source name;
Col.(2-5): the fitting results for $\rm CO($J$=1-0)$; Col.(6-9): the fitting results for $\rm CO($J$=2-1)$.}
\begin{tabular}{ccccccccccc}
\hline\hline \multicolumn{1}{c}{Source}
  & \multicolumn{8}{c}{Fitting results}\\
  & \multicolumn{4}{c}{$\rm CO(J=1-0)$}
  & \multicolumn{4}{c}{$\rm CO(J=2-1)$}\\
  \hline
  & Sensitivity & FWHM & Shift & Integral flux & Sensitivity & FWHM & Shift & Integial flux \\
  \hline
  & mk & \kms & \kms & Jy \kms &  mk & \kms & \kms & Jy \kms \\
   (1) & (2)  & (3)  & (4)  & (5)  & (6)   & (7)   & (8)   & (9) \\
\hline
PGC 32543  & 6.8$\pm$1.1 & 81.5$\pm$15.3               &  12.9$\pm$6.5 & 3.8$\pm$0.9  & 11.3$\pm$1.2 & 107.2$\pm$12.9  & 8.1$\pm$5.5 & 11.3$\pm$1.8  \\

PGC 35225 & 12.4$\pm$1.6 & 16.3$\pm$2.4    & -5.7$\pm$1.0 & 1.4$\pm$0.3   & 21.8$\pm$1.5 & 18.8$\pm$1.5  &-3.1$\pm$0.7 & 3.8$\pm$0.4  \\

PGC 36686 & 18.6$\pm$1.5 & 48.1$\pm$4.6    & 11.8$\pm$2.0 & 6.0$\pm$0.8   & 27.0$\pm$2.3 & 41.0$\pm$4.1 &  13.9$\pm$1.7 & 10.3$\pm$1.4  \\

PGC 44685 & 7.1$\pm$1.3 & 48.2$\pm$10.2    & 8.8$\pm$4.3 & 2.3$\pm$0.6   & 10.2$\pm$1.5 & 29.3$\pm$4.8  & 5.1$\pm$2.1 & 2.8$\pm$0.6  \\
\hline
\end{tabular}
\label{tab2} \centering
\end{table*}

\begin{figure*}
\centering
\includegraphics[angle=0,width=0.4\textwidth]{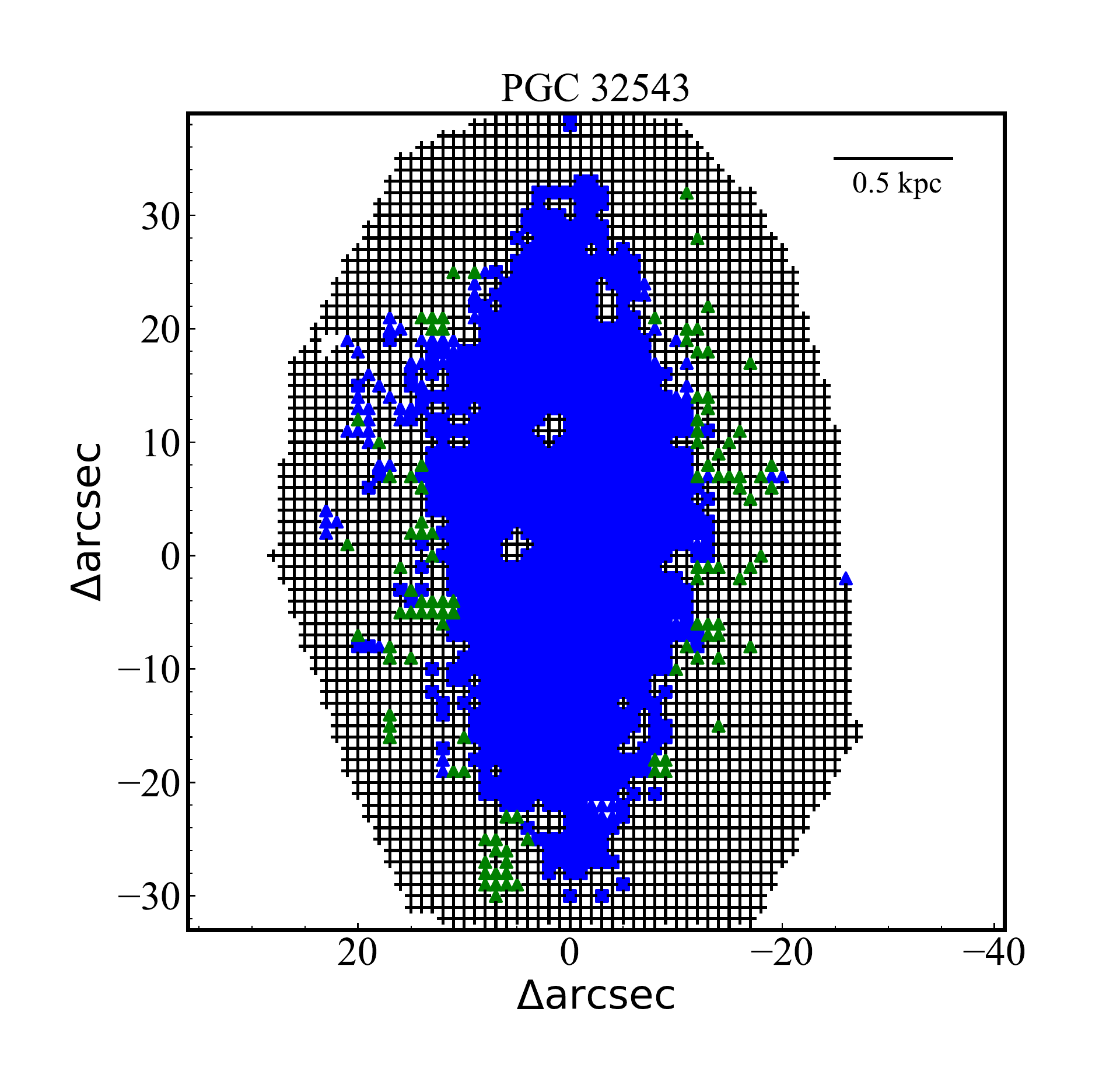}
\includegraphics[angle=0,width=0.4\textwidth]{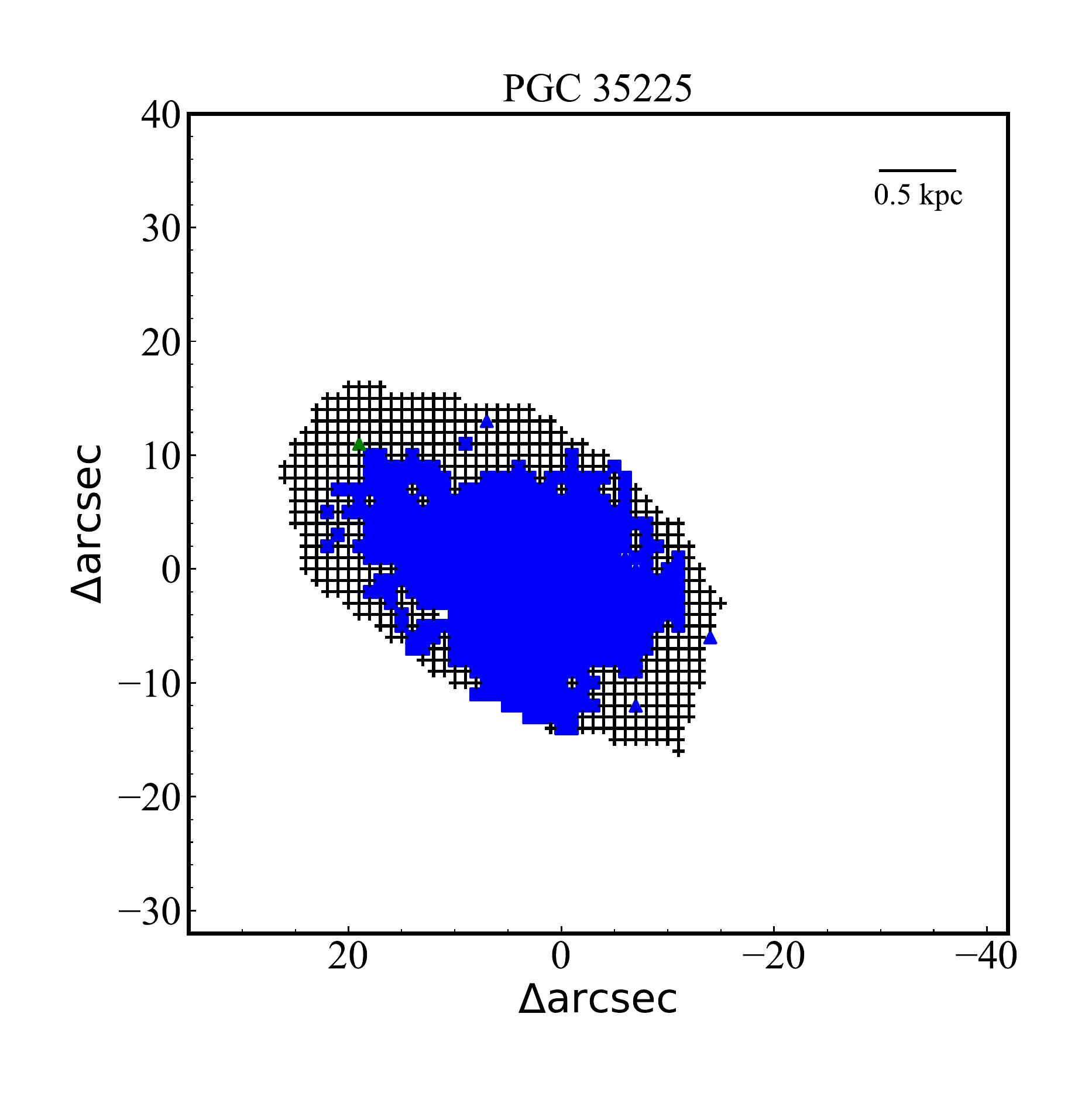}
\includegraphics[angle=0,width=0.4\textwidth]{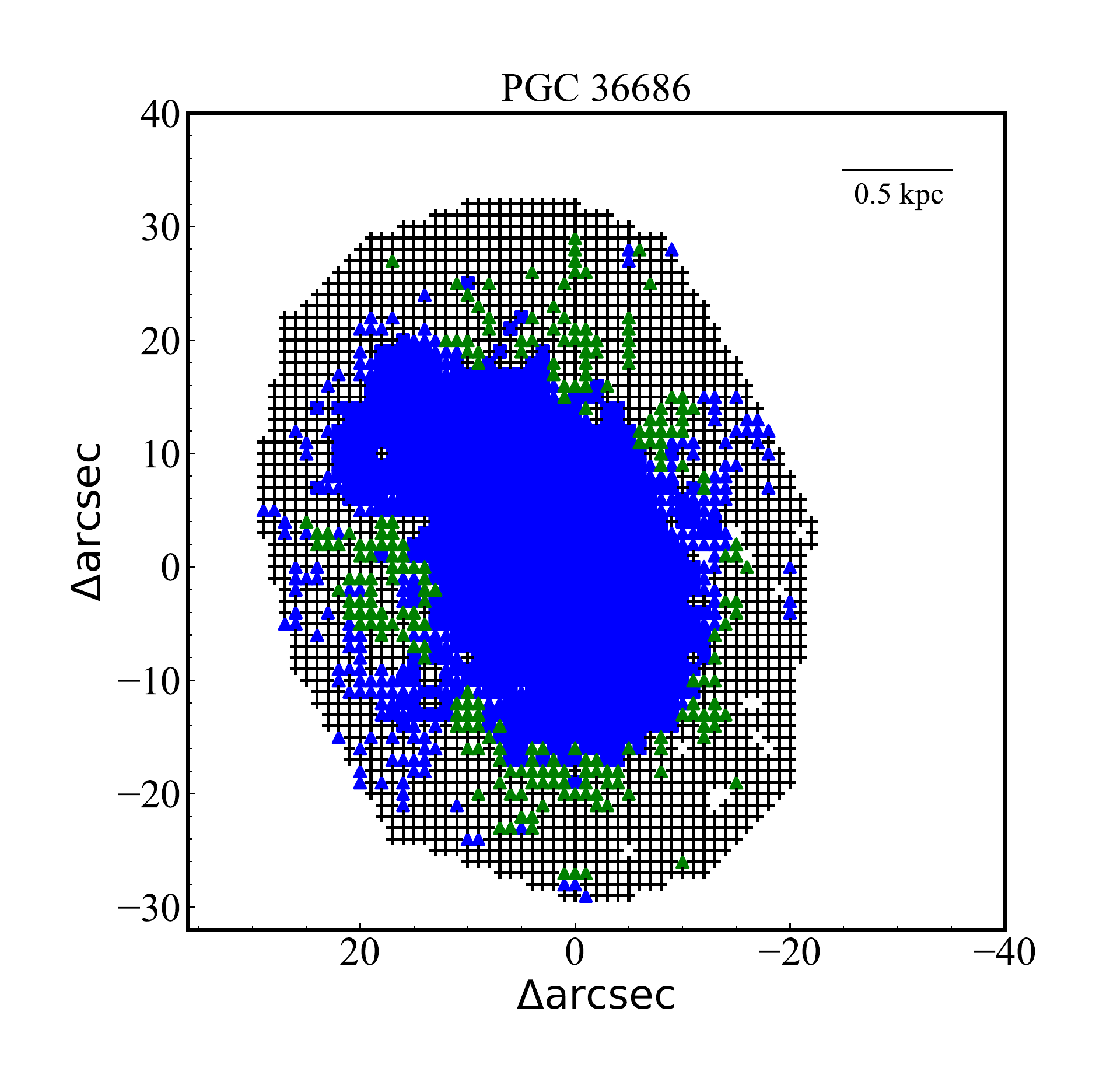}
\includegraphics[angle=0,width=0.4\textwidth]{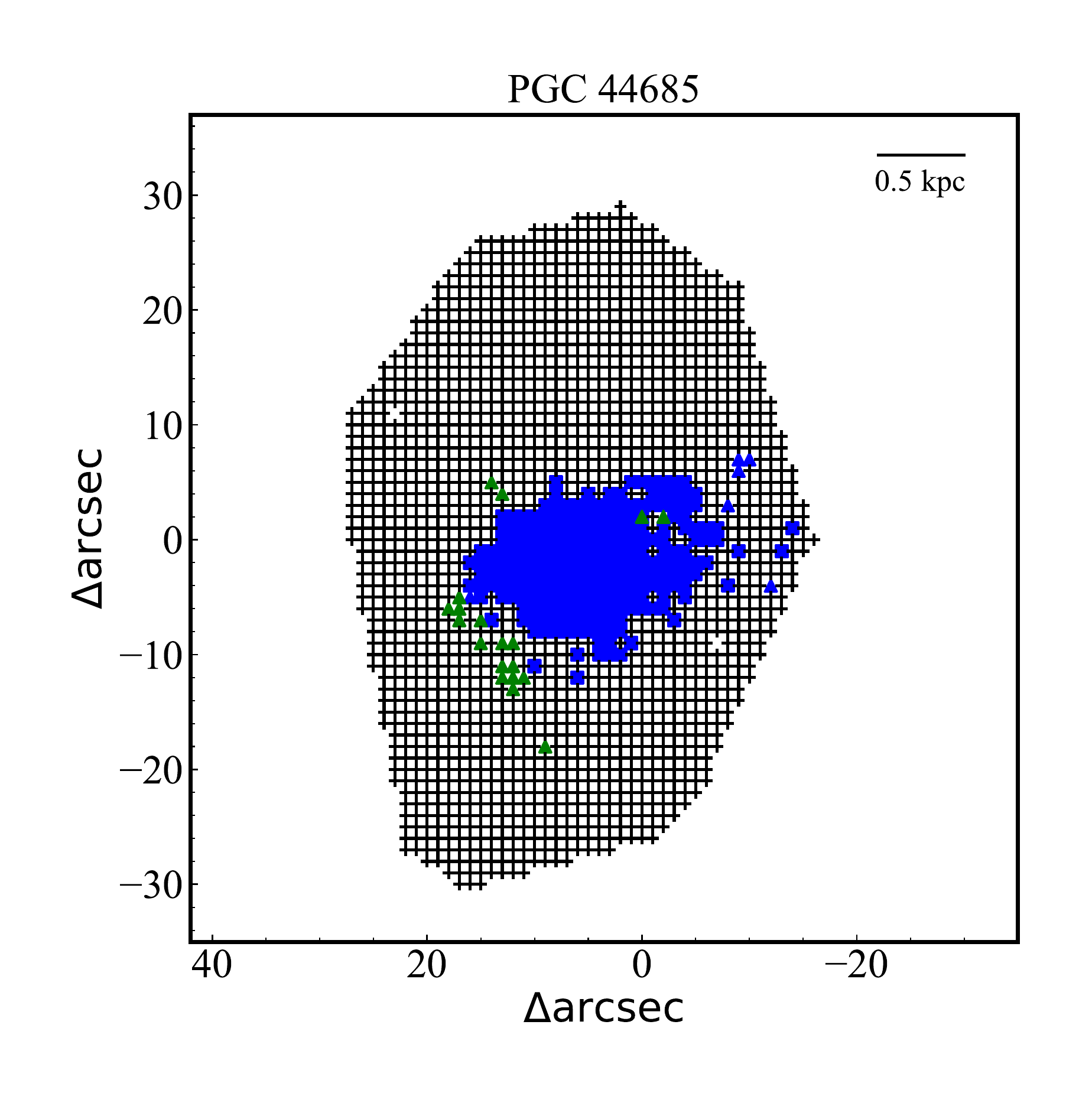}
\caption{The spatial distribution of ionization mechanism. The star formation and composite regions are marked as blue and green, respectively. The physical scale corresponding to each arcsec is marked in the upper right corner of each panel.}
\label{f2}
\end{figure*}

\subsection{CAHA 2D-Spectroscopic Observation}
\label{sec2.1}

The sources studied in this work have been covered by spatially resolved IFS observations using the PMAS/PPAK configuration mounted on the CAHA 3.5-m telescope in the Calar Alto observatory. The multi-aperture spectrometer equipped on the telescope has 382 fibers, each with a diameter of 2.7\arcsec, which produces a hexagonal field of view (FoV) of 78\arcsec $\times$ 73\arcsec\ for our sample. The FoV is sufficient to cover the entire galaxies up to two to three effective radii. CAHA observes the galaxies in two configurations (i.e., the V500 setup and the V1200 setup). 
The V500 setup covers the wavelength between 3745 and 7500 \AA\, which provides a low spectral resolution (R $\sim$ 850). The other setup covers the wavelength range from 3400 to 4840 \AA\, which provides a medium spectral resolution (R $\sim$ 1650).
The objects were observed by using a 3-pointing dithering scheme to obtain a filling factor of 100\%. The exposure time for per pointing was 900s. We used an upgraded python-based pipeline \citep{2015A&A...576A.135G, 2016A&A...594A..36S} to reduce the PPAK data. The reduction process are summarized as follows:
i) identification of the position of the spectra on the detector along the dispersion axis; ii) extraction of individual spectrum; iii) distortion correction of the extracted spectra; iv) wavelength calibration; v) fiber-to-fiber transmission correction; vi) flux-calibration; vii) subtraction of the sky; viii) datacube reconstruction; ix) and finally differential atmospheric correction. We recommend the readers to refer \cite{2013A&A...549A..87H}, \cite{2015A&A...576A.135G} and \cite{2016A&A...594A..36S} for more details of the reduction process. For the original data cubes, we correct the flux for Galactic extinction. The final spectra fully cover the wavelength range from 3700 to 7300 \AA.

\subsection{Spectroscopic Fitting and Parameter Calculation}
\label{sec2.2}

Firstly, we use full spectrum fitting method, pPXF \citep{2004PASP..116..138C, 2017MNRAS.466..798C}, to model the stellar continuum. pPXF adopts MILES simple stellar population templates \citep{2010MNRAS.404.1639V}, which assume \cite{1955ApJ...121..161S} IMF and \cite{2000ApJ...533..682C} dust extinction curve. The template library comprises 150 templates covering 25 population ages (from 0.06 to 15.85 Gyr) and 6 metallicities (i.e., log[M/H]=-1.71, -1.31, -0.71, -0.4, 0.0, 0.22). 
The stellar population fitting is applied to the V500 spectra and the emission lines are masked when modeling the stellar continuum.  
It is noted that we do not contain Active Galactic Nucleus in the templates considering these S0 galaxies were not identified as Active Galactic Nucleus through the BPT diagram in \cite{2016ApJ...831...63X}.

Next, we subtract the stellar continuum from the original spectra to obtain the ionized gas emission lines. For the emission lines, we fit them spaxel by spaxel
using single Gaussian models and the width of the Gaussian were tied for ions of the same element.

Considering that the emission lines may be contaminated by the AGNs or shocks, we tested the ionization mechanism of the emission lines of these galaxies. Figure \ref{f2}
shows the spatial distribution of the ionization mechanism of spaxels according to the traditional BPT diagnostic diagram \citep{2001ApJ...556..121K} and the equivalent width of
H$\alpha$ \citep{2018MNRAS.474.3727L}. We find that almost all emission lines are ionized by star formation (blue points in Figure \ref{f2}), especially in the central regions of galaxies (i.e., star formation regions). Therefore, we ignore the effects from AGN and shocks on the calculation of SFR.

With the fitting of emission lines, we calculate the spatially resolved SFR using extinction-corrected H$\alpha$ luminosity. The extinction is estimated using the Balmer decrement by assuming \cite{1989ApJ...345..245C} extinction curve with case B condition for each spaxel. For the spaxels without reliable extinction measurements (i.e., without strong H$\beta$ emission line), we adopt a zero extinction value to the spaxels to estimate the lower limit of SFR. Finally, the extinction-corrected SFR is computed by adopting a empirical formula given by \cite{1998ARA&A..36..189K}:
\begin{equation}
SFR(\msun yr^{-1})=7.9 \times 10^{-42} L(\rm H\alpha),
\label{e1}
\end{equation} 
where $L$(H$\alpha$) is the extinction-corrected H$\alpha$ luminosity. In order to calculate the total extinction-corrected SFR, we stack all spaxels and fit the H$\alpha$ luminosity with the continuum-subtracted emission line spectrum.

In order to obtain the surface mass density of atomic gas ($\Sigma_{\rm HI}$), we adopt a method based on gas-phase metallicity to estimate $\Sigma_{\rm HI}$ (see Table \ref{tab1}). 
The gas-phase metallicities of the four S0 galaxies are derived based on a 
Bayesian method\footnote{http://users.obs.carnegiescience.edu/gblancm/izi} \citep{2015ApJ...798...99B}, which assumes the \cite{2010AJ....139..712L} photoionization model. This method makes the probability density functions of gas-phase metallicity and ionization available after inputting a set of strong nebular emission lines. 
We use the stacked spectra that is used to estimate the total SFR to compute the gas-phase metallicity, which is used to derive $\Sigma_{\rm HI}$ according to the empirical formula of \cite{2018ApJ...862..110S}. We also compute the spatially resolved metallicity using a set of strong nebular emission lines of each spaxel.


\subsection{IRAM 30-m Observation and Data Reduction}
\label{sec2.3}

Spectra of the CO($J$=1-0) and CO($J$=2-1) emission lines in the four galaxies were observed in 2020 January with the IRAM 30-m telescope (Project 199-19. PI: Xue Ge). The full width at half maximum of the main beam of IRAM telescope is approximately $\sim$ 22\arcsec and 11\arcsec, respectively, in the two transitions of CO.

The sources can be observed with Eight MIxer Receiver (EMIR) using the tracked PSW (Position switching)/WSW (Wobbler switching) observing mode and the dual polarization measurements were obtained simultaneously in both CO($J$=1-0) and CO($J$=2-1). EMIR has four different bands (i.e., E090, E150, E230 and E330) and each represents a main atmospheric window in the millimetre.
Each of these bands has four intermediate frequencies outputs (i.e., 2 polarizations and 2 sidebands) and each of them covers 8 GHz bandpass. All the bandpass are split into two blocks equally denoted by I and O (inner and outer) and sent to the spectrometers by the use of in total 8 coaxial cables. 
The sources were observed using the wobbler switching.
We choose the Fast Fourier Transform Spectrometers (FTS) that can be connected to EMIR to obtain the emission lines of CO. The resolutions of CO($J$=1-0) and CO($J$=2-1) provided by FTS are 1.3 and 2.6 \kms wide, respectively. We use the closed and bright planets to check the telescope pointing and focus about every 1.5-2 hr.
The system temperatures were typically 200–400 K for both CO($J$=1-0) and CO($J$=2-1).

The spectra are reduced by the CLASS, a module of the GILDAS software package. All scans are summed and then averaged. The averaged spectra are smoothed to a typical resolutions of 10.0 \kms for both CO($J$=1-0) and CO($J$=2-1). 
Two narrow windows on both sides of the emission line are selected as the fitting windows of linear baselines and a first-order baseline is subtracted from the original spectra. 
For the baseline-subtracted emission-line spectra, we adopt the Gaussian model to fit CO($J$=1-0) and CO($J$=2-1). 

We find that all the four star-forming S0 galaxies show a significant CO($J$=1-0) and CO($J$=2-1) lines at a confidence level larger than 3$\sigma$. We convert the line intensities from main beam temperature $T\rm_{mb}$ (K) to flux density S (Jy) by adopting $S/T\rm_{mb} \sim$ 6.3 and 8.7 Jy K$^{-1}$ at 110 GHz and 235 GHz, respectively. The luminosities of CO lines are calculated using the standard equation from \cite{2005ARA&A..43..677S}: 
\begin{equation}
L\rm_{CO} = 3.25 \times 10^7 \it S\rm_{CO}\Delta \it V \nu\rm_{obs}^{-2} \it D\rm_L^2 (1+\it z)\rm^{-3}
\end{equation}
where $L\rm_{CO}$ is the luminosity in K km s$^{-1}$ pc$^2$ and the CO flux $S\rm_{CO}\Delta \it V$ obtained with single Gaussian model, is in Jy km s$^{-1}$. The observing frequency $\nu_{\rm obs}$ and the luminosity distance $D\rm_L$ are in GHz and Mpc, respectively. 
The conversion factor of the Galaxy between CO and H$_2$, $\alpha\rm{_{CO(1-0)}} = 4.35~\rm M_{\odot}\, \rm{pc^{-2}\, (K\,km\,s^{-1})^{-1}}$, corresponding to $X\rm_{CO} = 2 \times 10^{20}~cm^{-2}$ (K km s$^{-1})^{-1}$ \citep{2013ARA&A..51..207B} is adopted to convert the CO luminosity to the molecular gas mass.
Detailed information of the CO observations and the fitting results are summarized in Table~\ref{tab1} and Table~\ref{tab2}.



\section{Results}
\label{sec3}

\subsection{The Properties of Star Formation}
\label{sec3.1}

\begin{figure*}
\centering
\includegraphics[angle=0,width=0.41\textwidth]{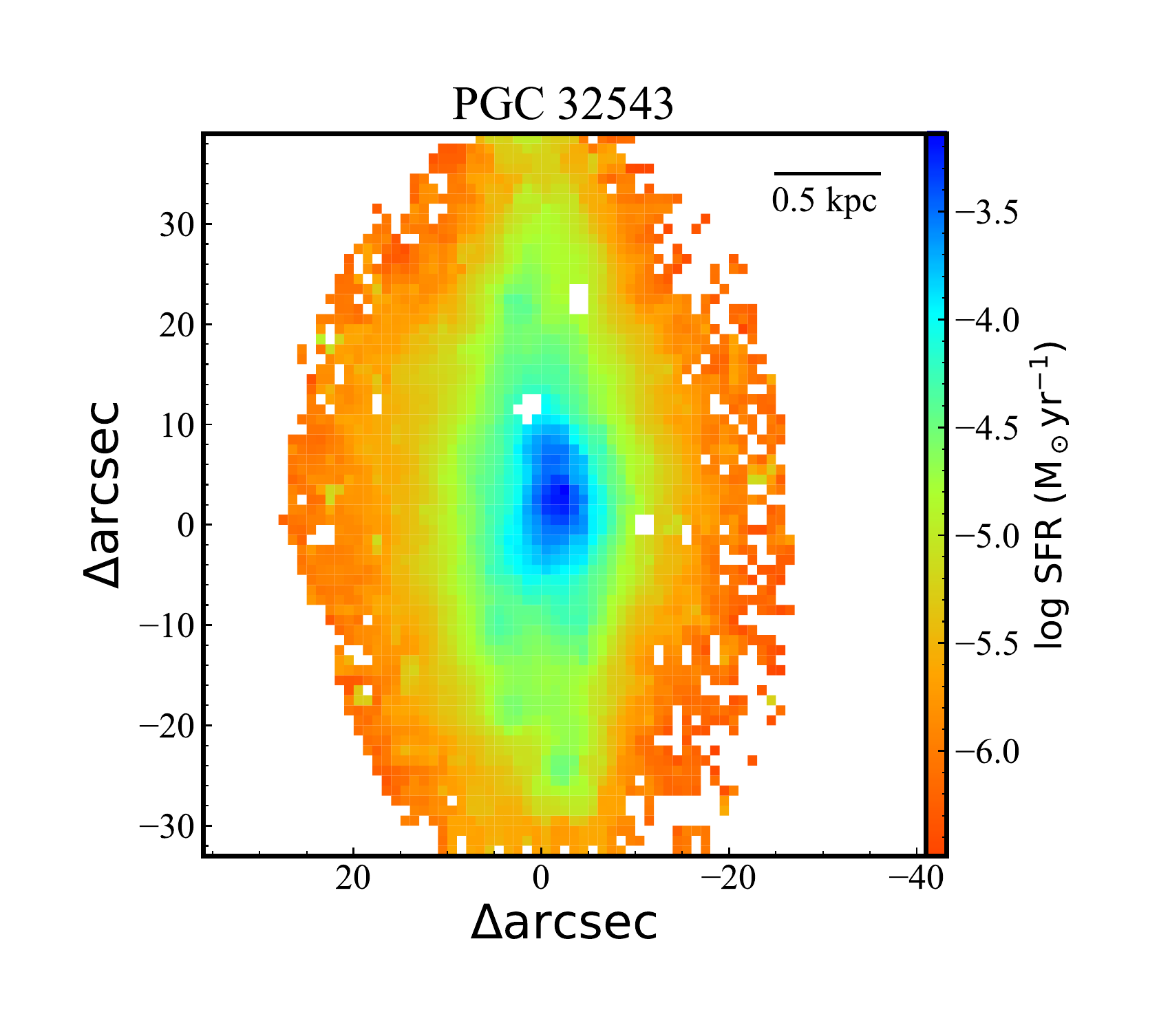}
\includegraphics[angle=0,width=0.41\textwidth]{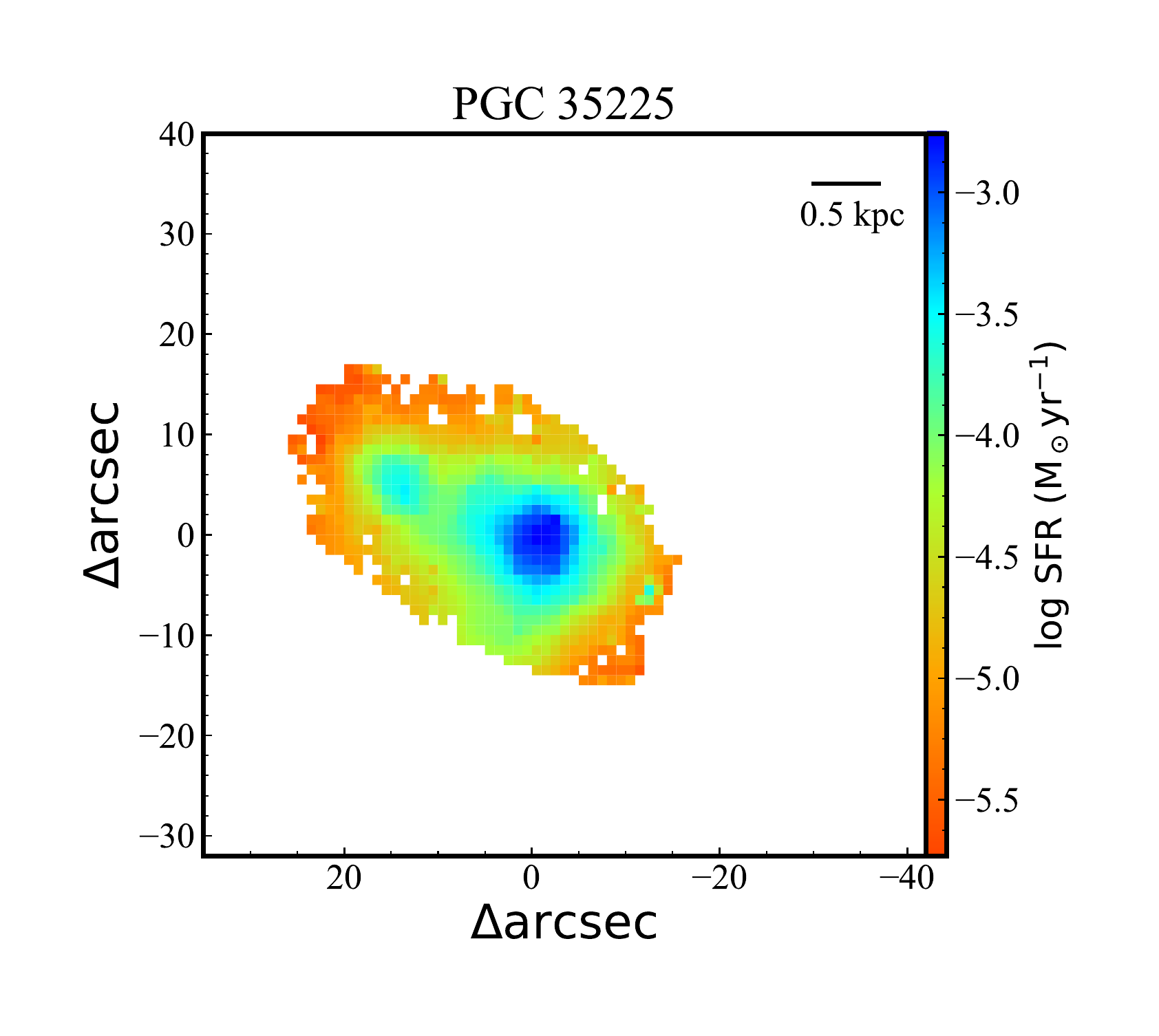}
\includegraphics[angle=0,width=0.41\textwidth]{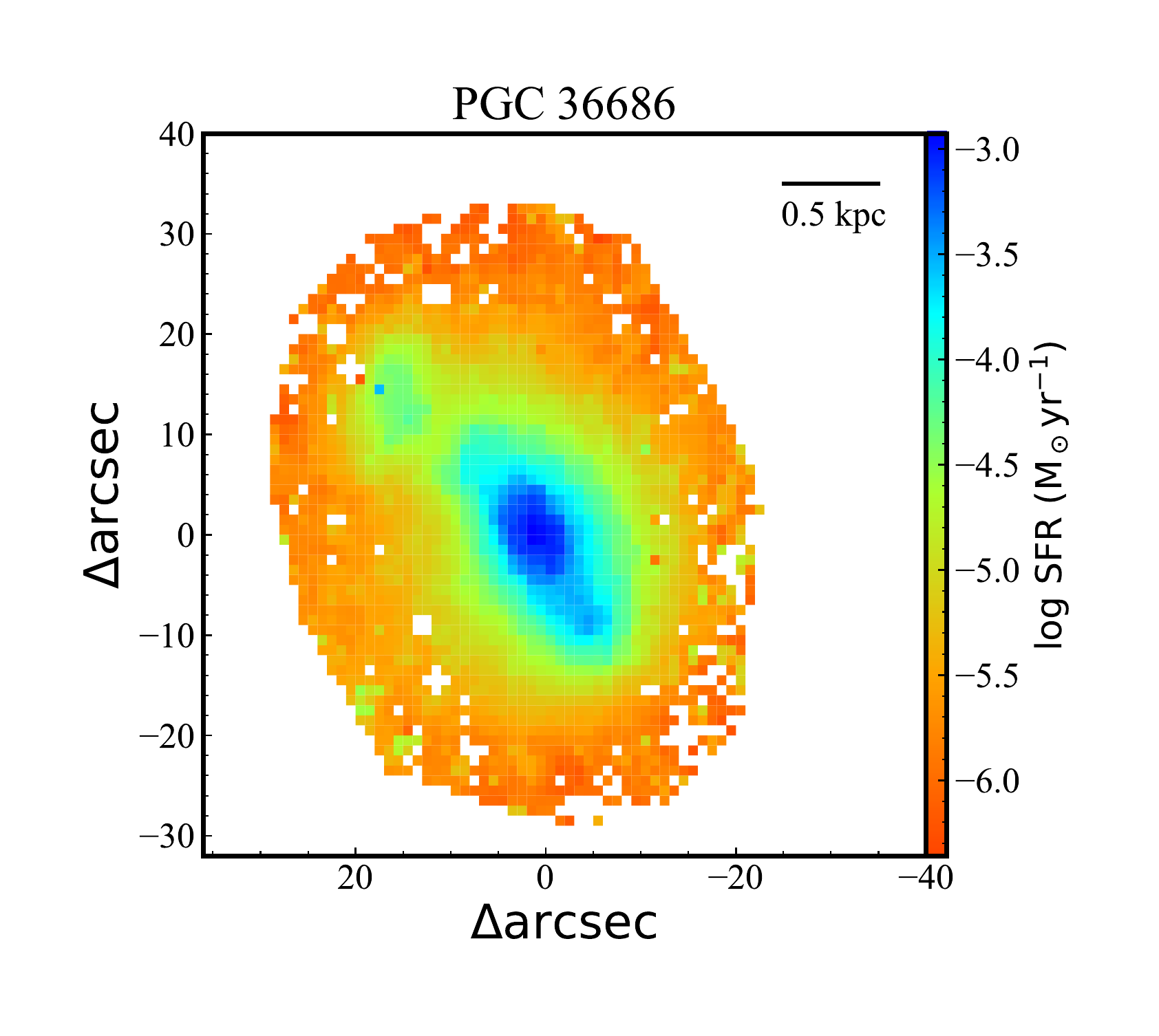}
\includegraphics[angle=0,width=0.41\textwidth]{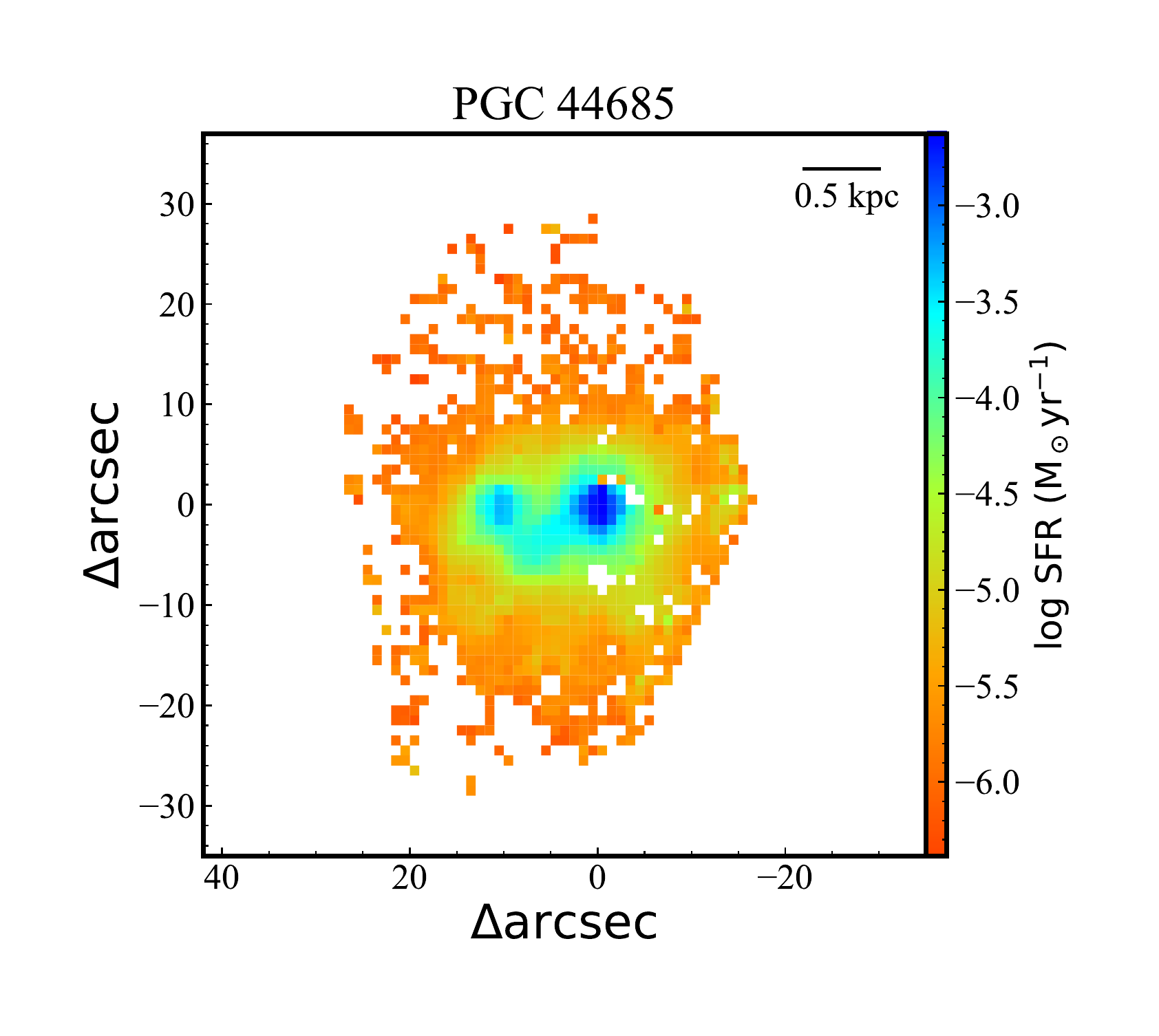}
\caption{Spatially resolved SFR for PGC 32543, PGC 35225, PGC 36686 and PGC 44685, respectively. The physical scale corresponding to each arcsec is marked in the upper right corner of each panel.}
\label{f3}
\end{figure*}

\begin{figure*}
\centering
\includegraphics[angle=0,width=0.6\textwidth]{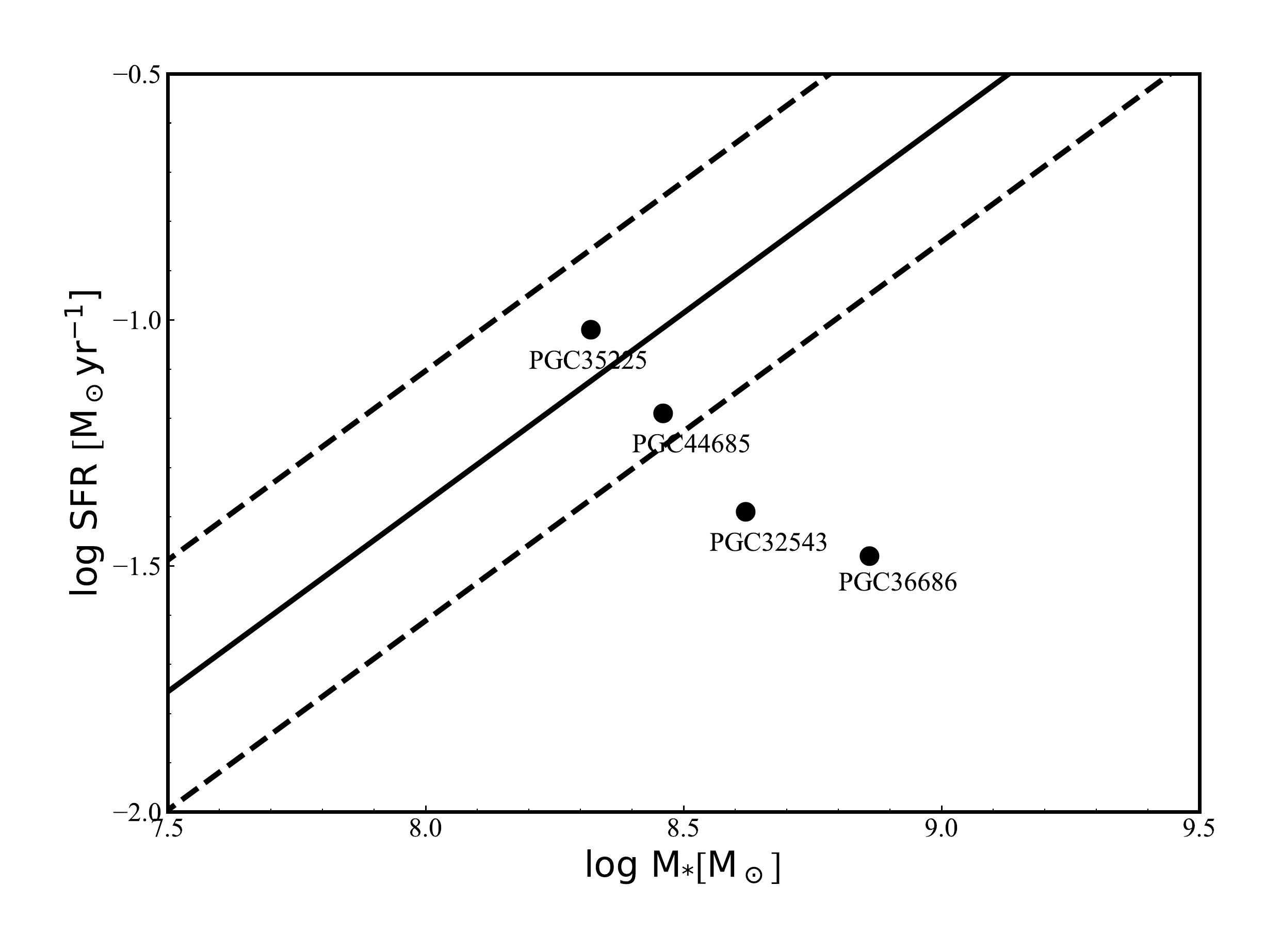}
\caption{Star formation rate as a function of stellar mass. The black solid line indicating the star-forming main sequence relation is adopted from \citet{2007A&A...468...33E} and the dashed lines represent the 68\% dispersion.}
\label{f4}
\end{figure*}

\begin{figure*}
\centering
\includegraphics[angle=0,width=0.5\textwidth]{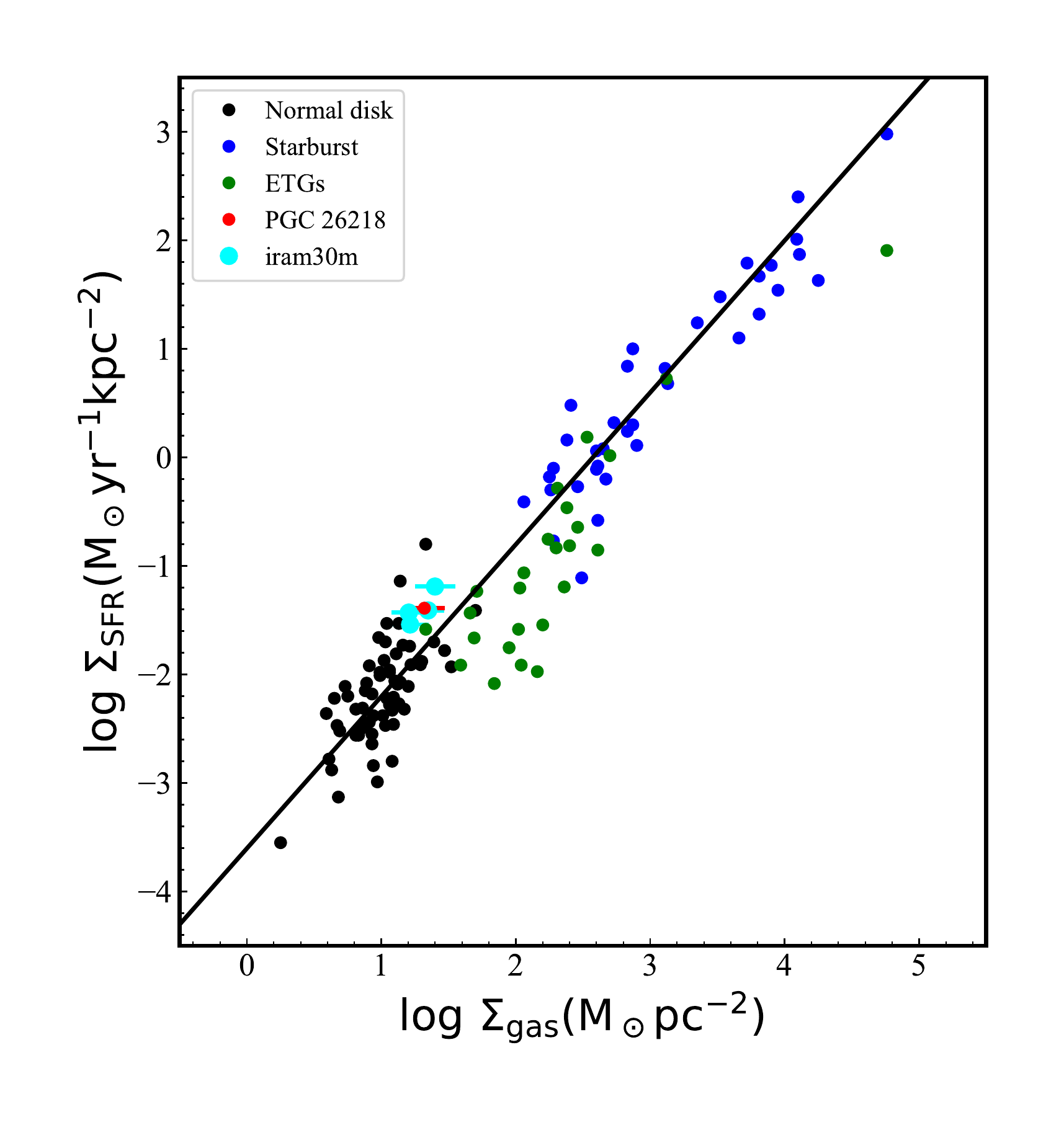}
\caption{$\log\ \Sigma_{\rm gas}$ vs $\log\ \Sigma_{\rm SFR}$ integrated over the beamsize of IRAM 30 m telescope ($\sim$ 22\arcsec) for CO($J$=1-0). The objects with spatially resolved CO($J$=1–0) detections (green points) are from \citet{2014MNRAS.444.3427D}, while the normal (black) and starburst (blue) galaxies are form \citet{1998ApJ...498..541K}.
The solid line represents the best fit for the \citet{1998ApJ...498..541K} sample. The red point indicates a star-forming S0 galaxy, which was studied by \citet{2020ApJ...889..132G}. The sources observed by IRAM 30 m telescope are marked in cyan.}
\label{f5}
\end{figure*}

In Figure \ref{f3}, we show the spatially resolved SFR of the four star-forming S0 galaxies. 
We find that there is no strong star formation in the outer region, while star formation mainly occurs in the center. The positions where the SFR is high correspond to the bright knots in Figure \ref{f1}. Such a pattern of star formation is consistent with the scenario where the star formation at the outskirts stops first, then at the central region (i.e., outside-in quenching mode). 
Although these S0 galaxies are forming stars, we can see from Figure \ref{f4} that some of them (e.g., PGC 32543 and PGC 36686) still deviate from the star forming main sequence relation \citep{2007A&A...468...33E, 2007ApJS..173..267S}. We suggest that it is the lack of star formation at the outskirts that causes them to deviate from the relation.
Another characteristic of star formation can be described by Kennicutt-Schmidt (K-S) law \citep{1959ApJ...129..243S, 1998ApJ...498..541K}, a relationship between the surface density of gas ($\Sigma_{\rm gas}$) and the surface density of SFR ($\Sigma_{\rm SFR}$). Without the maps of the objects, we do not know the detailed distributions of the molecular gas. So we calculated $\Sigma_{\rm gas}$ and $\Sigma_{\rm SFR}$ of each source using the star-forming spaxels (i.e., the ratios of gas mass and SFR to the area of the star-forming region). We find in Figure \ref{f5} that 
although ETGs (green points, most of them are S0 galaxies) deviate from K-S law, our sources all follow the relation. 

Many studies have shown that there is an inverse correlation between $\rm \alpha_{CO}$ and metallicity \citep{2008ApJ...686..948B, 2011ApJ...737...12L, 2013ApJ...777....5S} because metallicity traces the total dust content that can shield CO from Ultraviolet radiation. In this work, we adopt the $\rm \alpha_{CO}$ of the Galaxy. However, considering that our sources are low mass/metallicity, this assumption may underestimate the molecular gas mass. In order to determine how the molecular gas mass  changes, we estimate the $\rm \alpha_{CO}$ by using an empirical formula related to gas-phase metallicity and the distance off the main sequence given by \cite{2017MNRAS.470.4750A}. We find that the average $\rm \alpha_{CO}$ of these S0 galaxies is 8.79 (from 6.15 to 12.33), which leads to the $\Sigma_{\rm gas}$ being underestimated twice. Considering this conversion factor, which mainly depends on metallicity, our sources will move an average of 0.3 dex to the right in the K-S law (although this will lead these S0 galaxies to follow this law better). However, the formula from \cite{2017MNRAS.470.4750A} does not apply to galaxies merging or interacting with the environment (multiple star forming regions). Therefore, whether we really underestimate the molecular gas mass needs additional clues.
We note that the above results may have the selection effect, because these S0 galaxies were picked out via the visual morphologies and $B$-band magnitude \citep{2016ApJ...831...63X}.
\cite{2020ApJ...889..132G} studied the physical properties of a nearby star-forming S0 galaxy (red point in Figure \ref{f5}) via combining optical and millimeter data. 
Their results indicated that this S0 galaxy might have experienced a minor merger, which triggered the central star formation and resulted in the similar star formation efficiency to star-forming and starburst galaxies. Our result is consistent with that of \cite{2020ApJ...889..132G}. Based on the existence of multiple star forming regions, we suggest that the star formation of these S0 galaxies is probably related to a merger event (see Section \ref{sec4} for a detailed discussion of this issue).

\subsection{Kinematics of Star and Ionized Gas}
\label{sec3.2}

\begin{figure*}
\centering
\includegraphics[angle=0,width=0.8\textwidth]{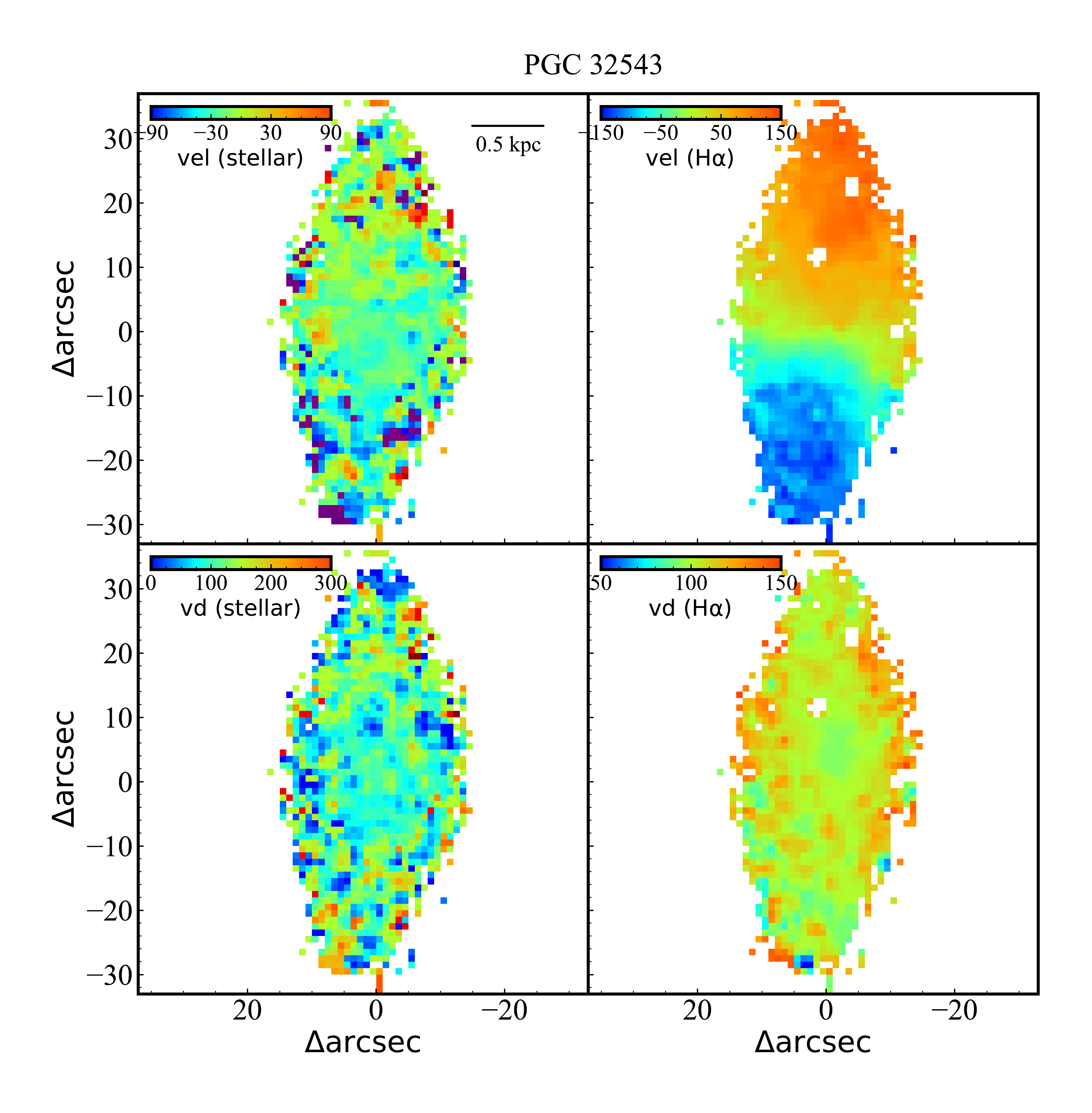}
\caption{The distributions of kinematics for PGC 32543. The component of velocity is labeled in the upper left corner of each panel. Note that we only select spaxels with S/N higher than 10 to measure the stellar velocity. The physical scale corresponding to each arcsec is marked in the upper right corner of the first panel.}
\label{f6}
\end{figure*}

\begin{figure*}
\centering
\includegraphics[angle=0,width=0.8\textwidth]{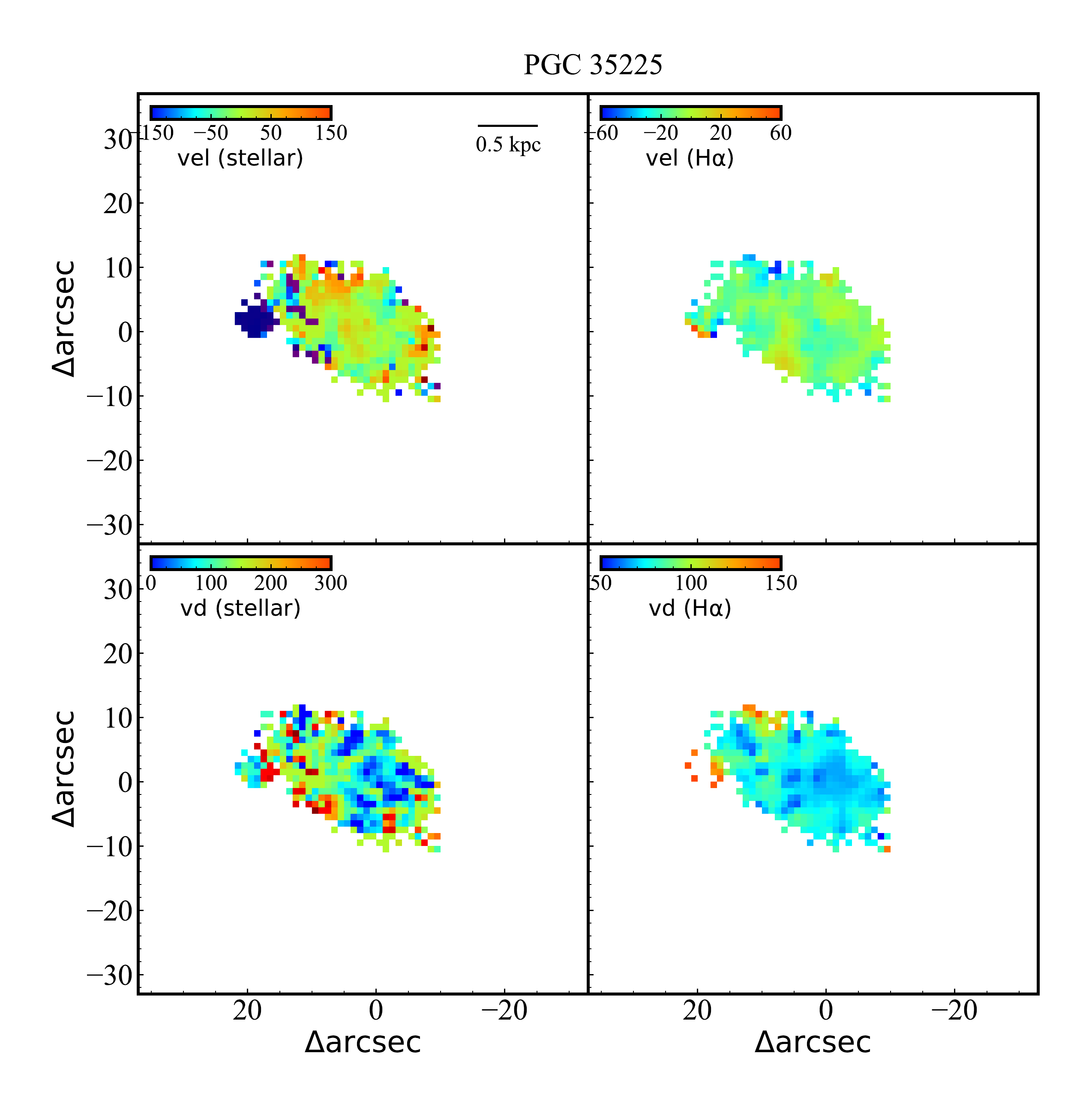}
\caption{As for Figure \ref{f6}, but for PGC 35225.}
\label{f7}
\end{figure*}

\begin{figure*}
\centering
\includegraphics[angle=0,width=0.8\textwidth]{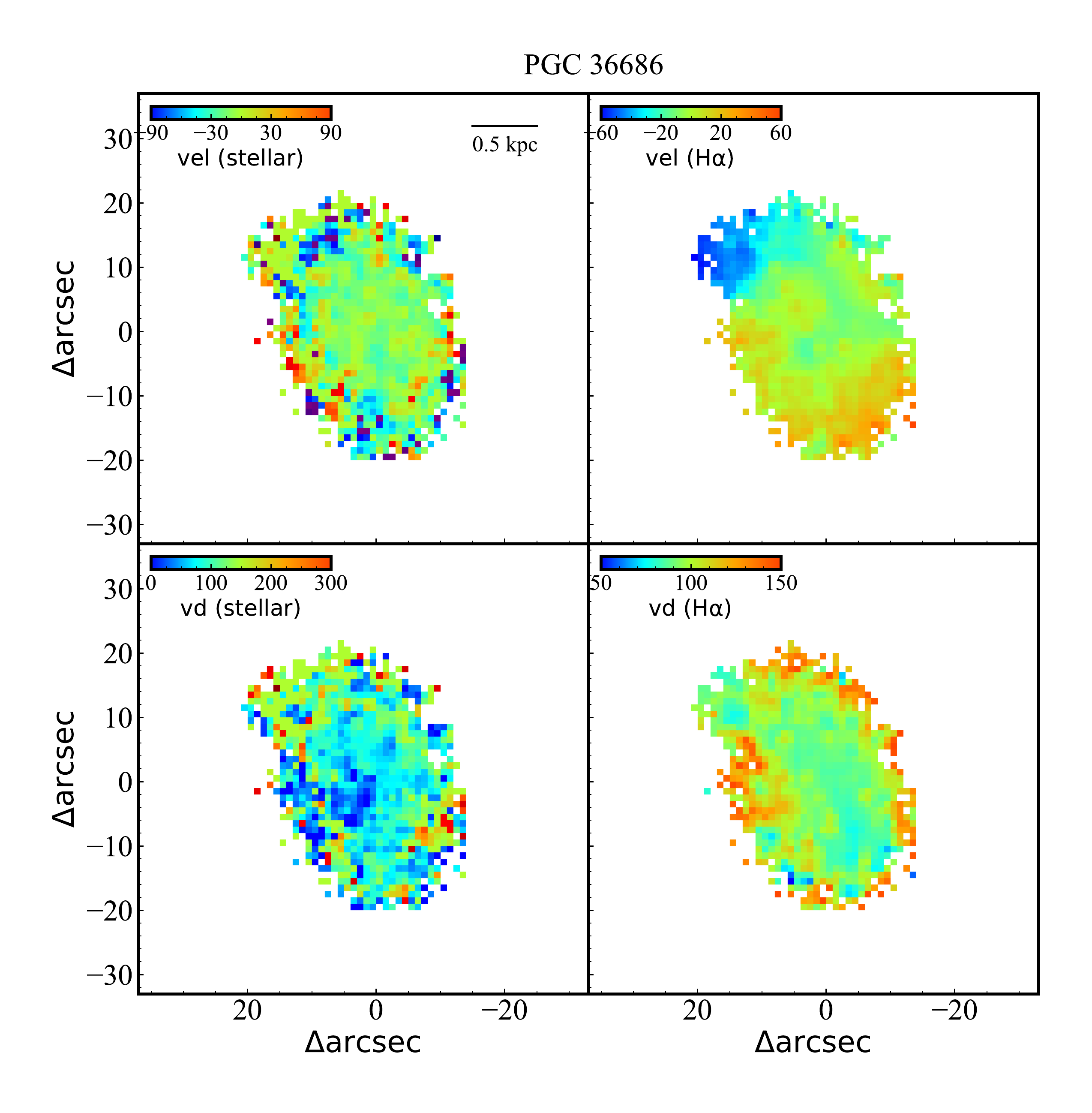}
\caption{As for Figure \ref{f6}, but for PGC 36686.}
\label{f8}
\end{figure*}

\begin{figure*}
\centering
\includegraphics[angle=0,width=0.8\textwidth]{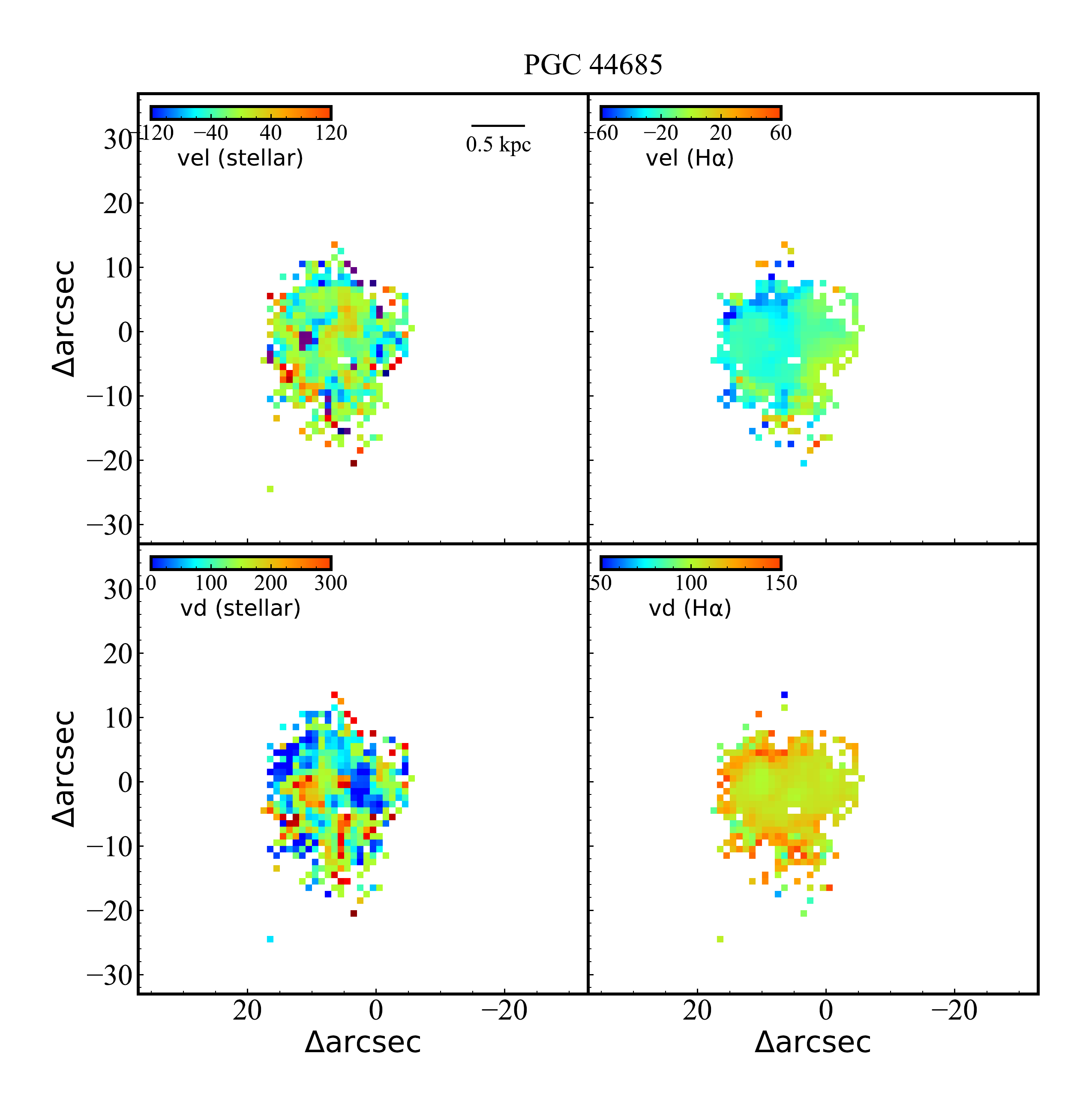}
\caption{As for Figure \ref{f6}, but for PGC 44685.}
\label{f9}
\end{figure*}

In Figure \ref{f6} - \ref{f9}, we compare the kinematics of stars and H$\alpha$ emitting ionised gas. It is noted that we only select the spaxels with signal-to-noise ratio (S/N) higher than 10, which would have reliable velocity measurements. We find that the velocity fields of star (the shift of the stellar absorption lines relative to systematic velocity) of all sources are disorganized, which indicates that these star-forming S0 galaxies do not show structures characteristic of a rotating disk. Similarly, we find that the velocity fields of ionized gas (the shift of emission line relative to systematic velocity) do not show regular rotating structure, except for PGC 32543.
We derive the axial ratios based on two-dimensional, single-component S\'ersic fit in $r$-band from SDSS website \footnote{https://www.sdss.org/dr16/manga/manga-target-selection/nsa/} and find that the axial ratios for PGC 32543, PGC 35225 and PGC 36686 are approximately 0.51. This shows that these sources are not entirely face-on galaxies. On the other hand, PGC 44685 does not display the rotated disk-shape structure, probably because it is a blue compact dwarf analysed in \cite{2013ApJ...764...44Z} or because it is a face-on galaxy based on the decomposition of bulge and disk \citep{2016ApJ...831...63X}.
In addition, we also find that the velocity dispersions of stars and ionized gas are low in the regions with high star formation, which indicates that these star-forming S0 galaxies still have significant rotation. It is can be expected if these systems are starburst dwarfs.

Based on the kinematics of stars and ionized gas, we calculate $\lambda_{R_{e}}$ defined as 
\begin{equation}
{\lambda_R}_e = \frac{\sum_{i=1}^{N_p} F_i R_i \left| V_i \right|}{\sum_{i=1}^{N_p} F_i R_i \sqrt{V_i^2+\sigma_i^2}} \, ,
\label{eq:sumLambda}
\end{equation}
within an effective radius and we simply take $1-q$ ($q$ is axial ratio) as the ellipticity $\epsilon$. In $\lambda_{R_{e}} - \epsilon$, we find that the values of these two parameters are (0.22, 0.47), (0.43, 0.50), (0.17, 0.48) and (0.40, 0.20) for PGC 32543, PGC 35225, PGC 36686 and PGC 44685, respectively. Slow and fast rotators have $\lambda_{R_{e}}$ lower and higher than $k_{FS} \times \sqrt{\epsilon}$, respectively, where $k_{FS} = 0.31$ for measurements made within an effective radius \citep{2011MNRAS.414..888E}. It is found that except for PGC 36686, other sources are fast rotators. However, contamination from various physical processes that can shape galactic morphologies, such as interaction and merging, gas accretion, environmental effects, secular evolution may lead to misjudgment. In particular, the study of the kinematics of dwarf ETGs still needs larger sample.
In general, the kinematic turbulence of star and ionized gas might imply that the fuels for the star formation might come from external environment.

\subsection{The Observation of Cold Gas}
\label{sec3.3}

\begin{figure*}
\centering
\includegraphics[angle=0,width=0.45\textwidth]{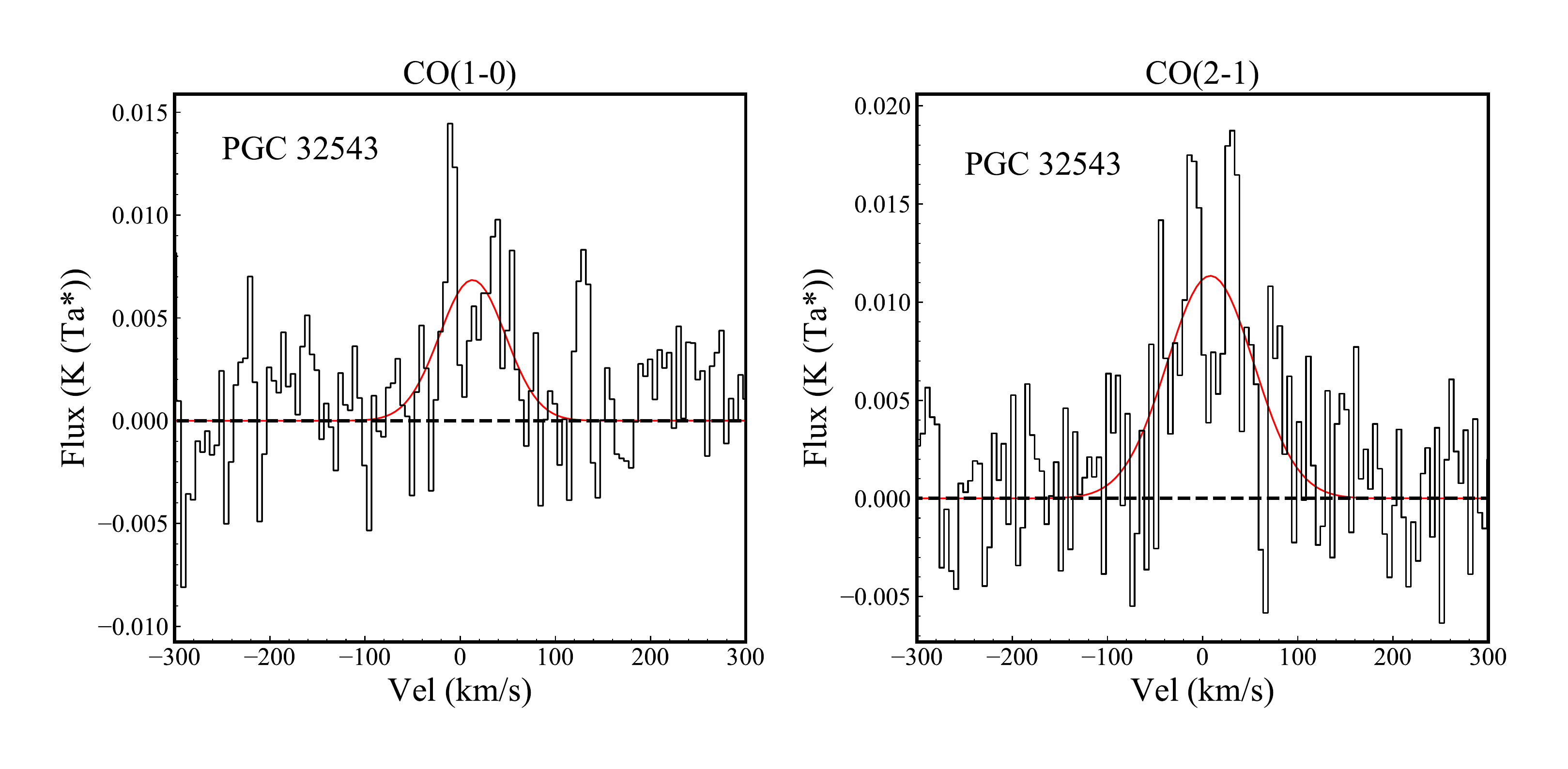}
\includegraphics[angle=0,width=0.45\textwidth]{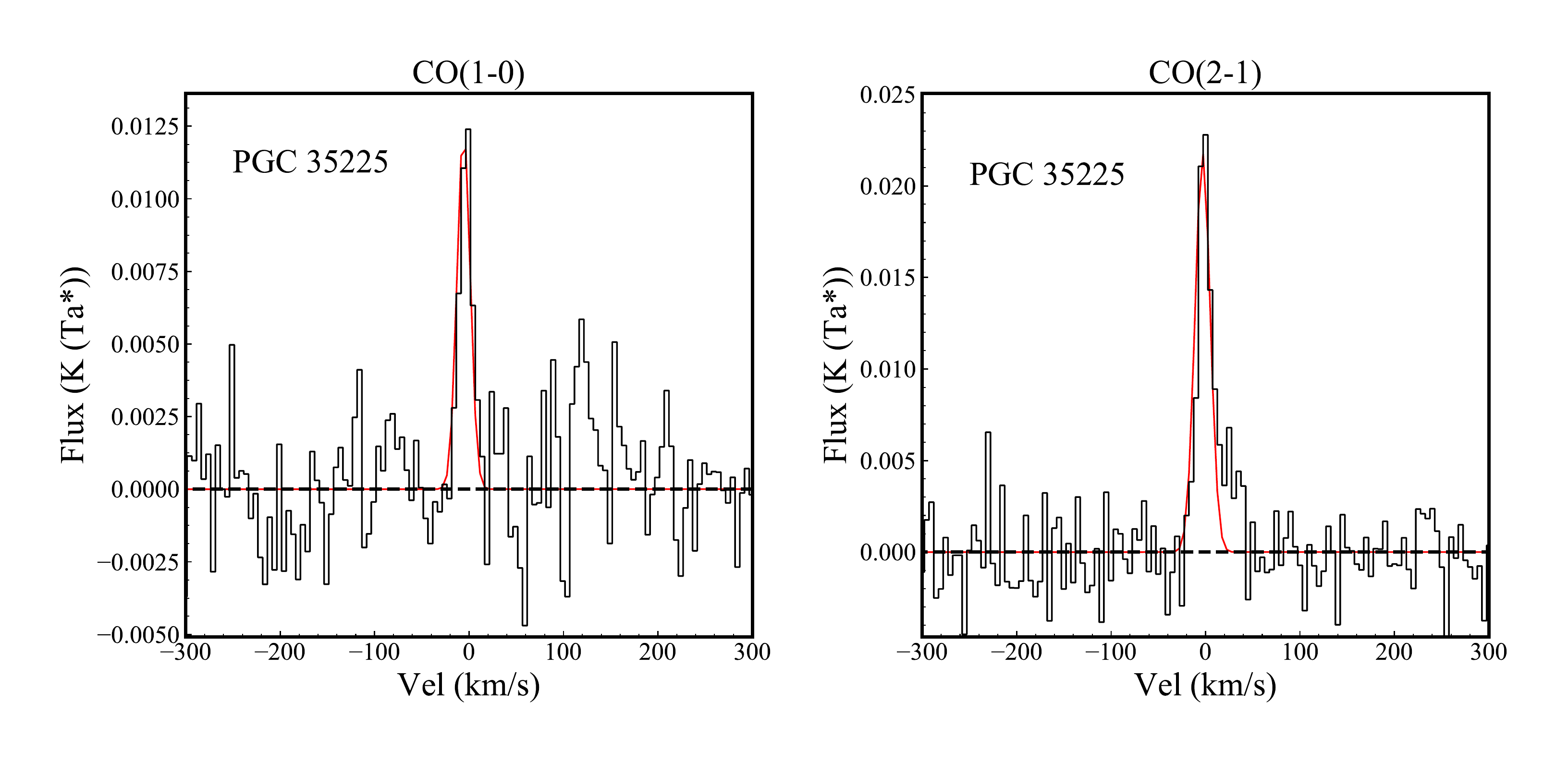}
\includegraphics[angle=0,width=0.45\textwidth]{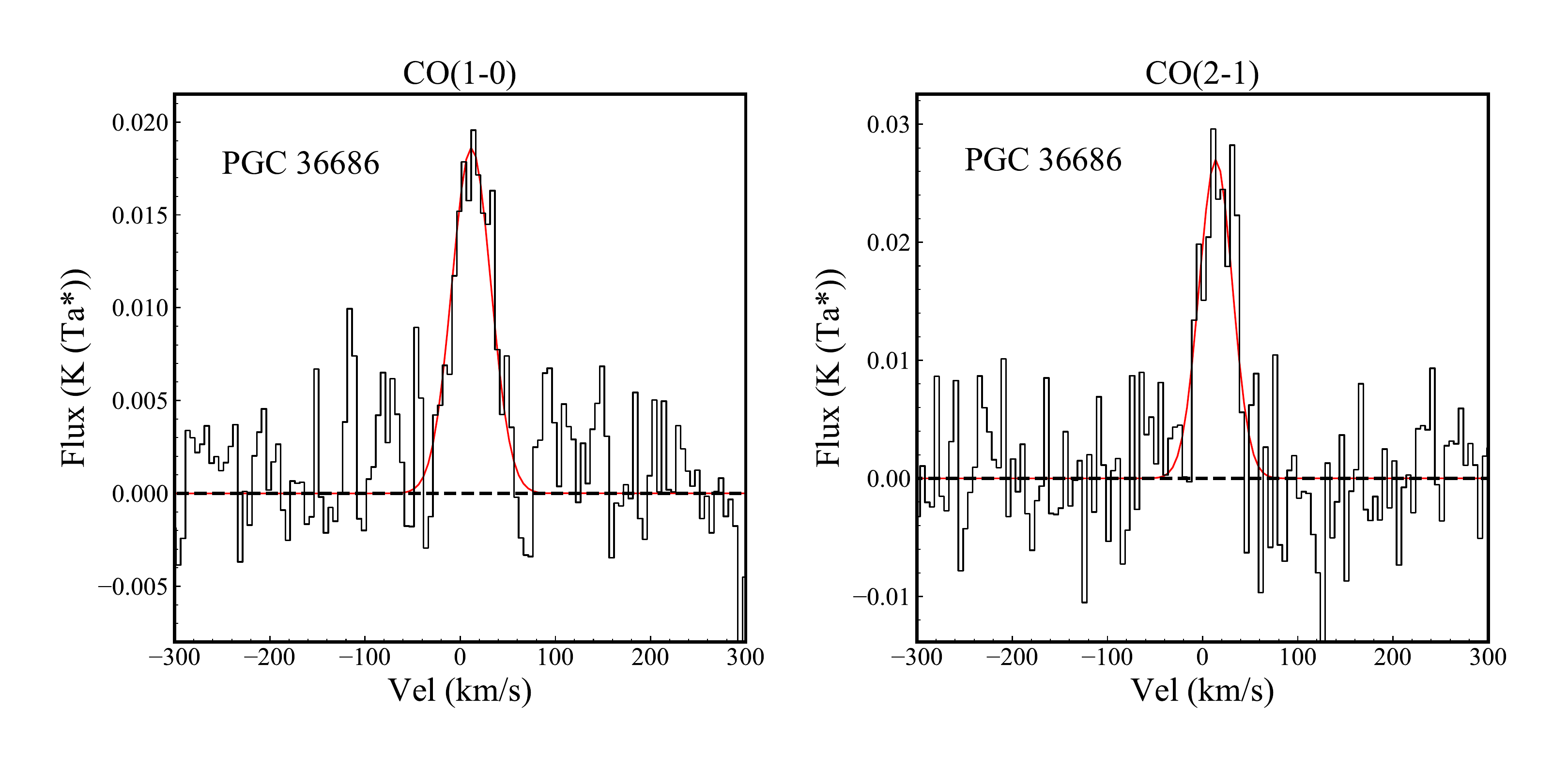}
\includegraphics[angle=0,width=0.45\textwidth]{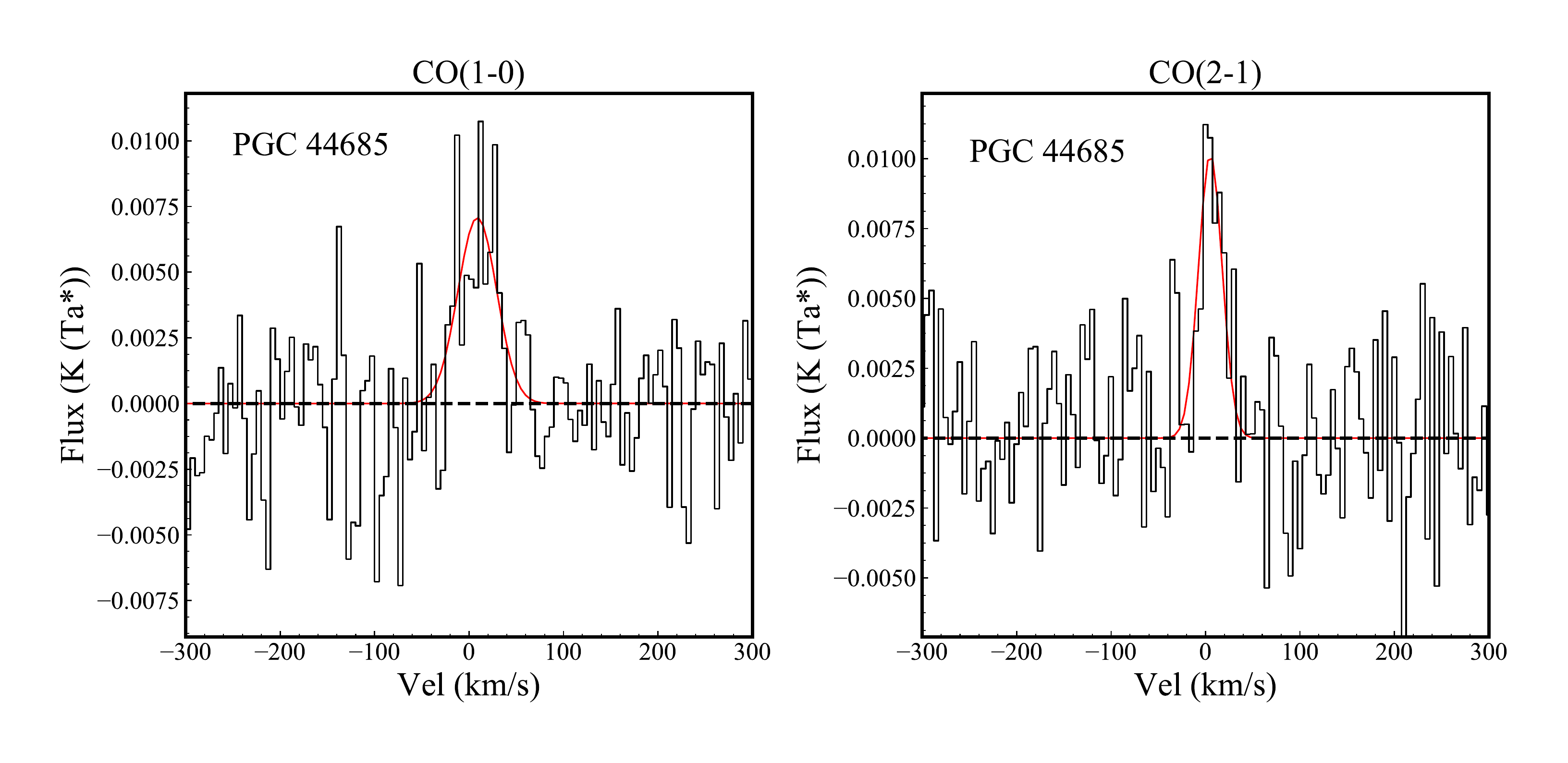}
\caption{The observed CO($J$=1-0) and CO($J$=2-1) spectra centred on systemic redshifts taken from SDSS for our sample. The black and red lines represent the original spectra and single Gaussian model, respectively.}
\label{f10}
\end{figure*}

From the best fitting of CO emission lines in Figure \ref{f10}, we note that the four S0 galaxies have masses of molecular gas less than $10^{7}~\rm M_{\odot}$. In addition, the fraction of molecular gas ($M_{\rm H_2}/M_{\ast}$) is the lowest 0.8\% for PGC 32543 and the highest 1.5\% for PGC 35225. This result is consistent with their positions in the main sequence relation (see Figure \ref{f4}). However, we find that the mass of the atomic gas is about 1-2 orders of magnitude higher than that of molecular gas and is of the order of the stellar mass. 
The molecular gas arises mostly from the stellar mass returned to the galaxy, while the atomic hydrogen is mainly accumulated from external sources (infall, captured dwarfs, etc.).
In this work, these dwarf S0 galaxies are difficult to form stars by using the molecular hydrogen returned by stars considering their low gravitational potential well. Therefore, the material for the star formation is more likely to come from external environment where the atomic hydrogen is easier to be obtained.
This is why the mass of atomic hydrogen is significantly greater than that of molecular hydrogen in our work.
These discoveries suggest that the star-forming materials of these galaxies may be derived from their rich atomic gas. The source of the rich gas may come from the merger events \citep{2015MNRAS.449.3503D} or the accretion of cold gas \citep{2013ApJ...770...62D}.

When the empirical formula from \cite{2018ApJ...862..110S} is used, the $\Sigma_{\rm HI}$ may be underestimated considering these S0 galaxies
may be accreting gas from the surroundings, therefore reduce the metallicity and thus the density of the atomic hydrogen. If in these cases, the $\Sigma_{\rm gas}$ should move to a larger region. The other factor  that may affect the estimation of $\Sigma_{\rm gas}$ is that we do not know how the gas is distributed. However, the resolution of CO($J$=1-0) produced by the IRAM-30 m telescope can basically cover the star forming regions of these dwarf galaxies. Therefore, it would not significantly change our results.

We also find that the FWHM of CO($J$=1-0) is $\lesssim$ 100 \kms, which might be caused by the concentration of gas in the center. The disturbed disks of stars and ionized gas may also support this result.
We utilize Tully-Fisher relation (i.e., $M_{\ast}$ vs FWHM) given by \cite{2016MNRAS.461.3494T} to estimate the FWHM of these S0 galaxies. However, the line widths of only two sources (PGC 32543 and PGC 44685) are consistent with our fitting results. The CO($J$=1-0) profile of PGC 35225 shows the asymmetry, and there seems to be absorption on the left side of the emission line. This non-gaussian in shape may cause it to deviate from the Tully-Fisher relation. It noted that the relation between the stellar mass and FWHM of CO($J$=1-0) from \cite{2016MNRAS.461.3494T} is based on the sample with $M_{\ast}\gtrsim9.5$. Whether the low-mass S0 galaxies conform to this relationship needs to be verified by larger samples.

The line ratios between CO($J$=2-1) and CO($J$=1-0) (i.e., $I_{21}/I_{10}$) can be used to investigate the physical properties of molecular gas, although the emission from
dense and diffuse surrounding molecular gas would complicate the interpretation. 
The line ratio is determined by the excitation temperature and the spatial distribution of the cold gas. When the CO gas is compacted relative to the beam, then CO ($J$=2-1) is stronger than CO ($J$=1-0). However, if the distribution of CO emission is not dense enough, then the intensity of CO ($J$=2-1) will be weakened more than that of CO ($J$=1-0), which will result in a ratio of less than 1. 
The values of $I_{21}/I_{10}$ for PGC 32543, PGC 35225, PGC 36686 and PGC 44685 are 74\%, 69\%, 43\%, and 30\%, respectively. 
This result might indicate that the cold gas traced by different CO emission lines either is very diffuse, or has significantly different excitation temperatures. 
Some previous studies \citep{1993ApJ...411L..71L, 1990A&A...239..125S, 1997A&A...323..727W} have shown that $0.5 \lesssim I_{21}/I_{10} \lesssim$ 1.0 imply that the gas is optically thick.
In addition, a study in $\rm ATLAS^{3D}$ has shown that the  integrated CO ($J$=2-1)/CO ($J$=1-0) of ETGs is almost between 1 and 4 \citep{2011MNRAS.414..940Y}. The difference between this work and \cite{2011MNRAS.414..940Y} may be either caused by the subthermal excitation or by the pointing errors which causes
the intensity of the CO($J$=2-1) line to be underestimated.
We find that object with high ratio of integrated intensity has low SFR (e.g., PGC 32543), while the object with low ratio has relatively high SFR (e.g., PGC 44685). Such a contradiction may be related to the telescope beams couple and the inevitable mixing of emission from dense and diffuse molecular gas. A larger sample is needed to investigate the relationship between the ratio of different transitions and the star-forming activity.

\section{Discussion}
\label{sec4}

In broad terms, galaxies could form via either the monolithic formation or the hierarchical accumulation. In the monolithic formation mechanism, a large amount of gas collapsed rapidly to form galaxies. The speed of collapsed gas  determines whether elliptical galaxies, S0 galaxies or spiral galaxies with spheroidal bulge are formed \citep{1987A&A...173...23A, 1988A&AS...75...93R, 1992MNRAS.254..601B, 1993ApJ...405..538B}. 
S0 galaxies are generally thought to have formed most of their stellar mass 10 Gyr ago, which is why they are quiescent at present.
This view has been reinforced by the paucity of atomic gas \citep{1986AJ.....91...23W} and star-forming regions \citep{1993AJ....106.1405P}. \cite{2003ApJ...584..260W} studied the relation between the total mass of atomic and molecular gas and blue luminosity of S0 galaxies and they found that S0 galaxies has lost almost 90\% of their gas.
However, the main disadvantage of this theory is that the massive protogalaxies that should be observed have not been observed at any reshift. In addition, S0 galaxies formed in this way should have a bulge component and have little star formation. In our sample, we do not find that these S0 galaxies have a large s\'ersic index. As \cite{2016ApJ...831...63X} pointed out, S0 galaxies with star formation have lower s\'ersic index than that without star formation.
The deficiencies in the monolithic formation have stimulated investigations of the hierarchical formation idea, where galaxies grew their stellar mass through a series of merger events \citep{1991ApJ...379...52W, 1993MNRAS.264..201K, 1996MNRAS.283.1361B}.
Clumps of different mass and angular momentum interact, which may lead to continuous star formation. At the same time, the evolution of galaxies will continue for a long time, and even at the present epoch we may see the relic of interaction (e.g., the multiple star-forming knots).

The study from \cite{2014MNRAS.437L..41K} presented that ETGs host about 14 per cent of the star formation budget which is caused by the minor merger events. It indicates that the minor mergers may play an important role in driving the star formation and leading perturbed stellar morphology . Based on a toy model, \cite{2016MNRAS.457..272D} found that a high gas-rich mergers rate may be required to explain more efficient star formation in local ETGs. Considering the major-merger rate is lower significantly than that of the minor-merger rate \citep{2011ApJ...742..103L}, therefore gas-rich minor mergers dominate the accretion of cold gas \citep{2011MNRAS.411.2148K}. For these star-forming S0 galaxies, they are morphologically disturbed (e.g., the star-forming knots) and have turbulent  kinematics, which supports the picture where the minor mergers can be a dominate factor for the star formation in the later morphological type.

\begin{figure*}
\centering
\includegraphics[angle=0,width=0.42\textwidth]{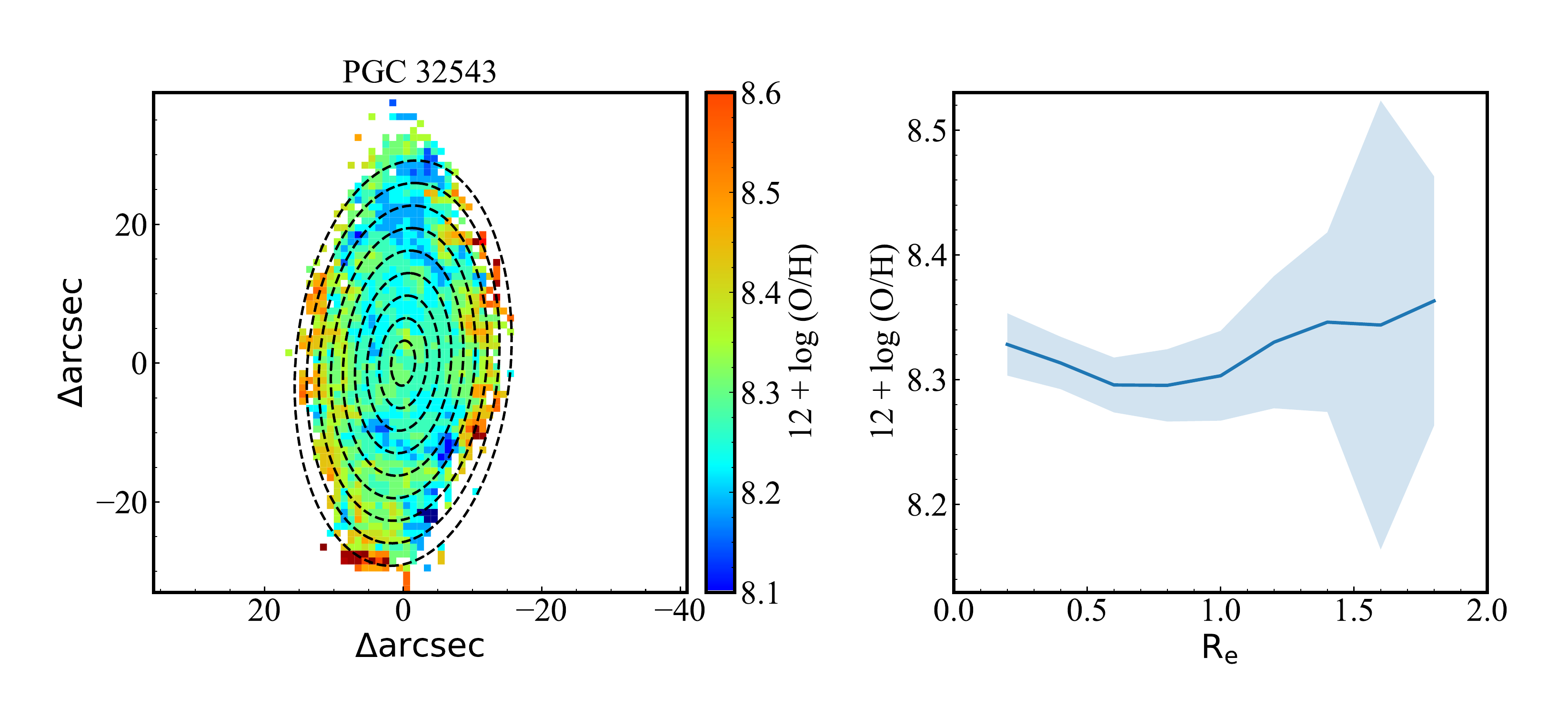}
\includegraphics[angle=0,width=0.42\textwidth]{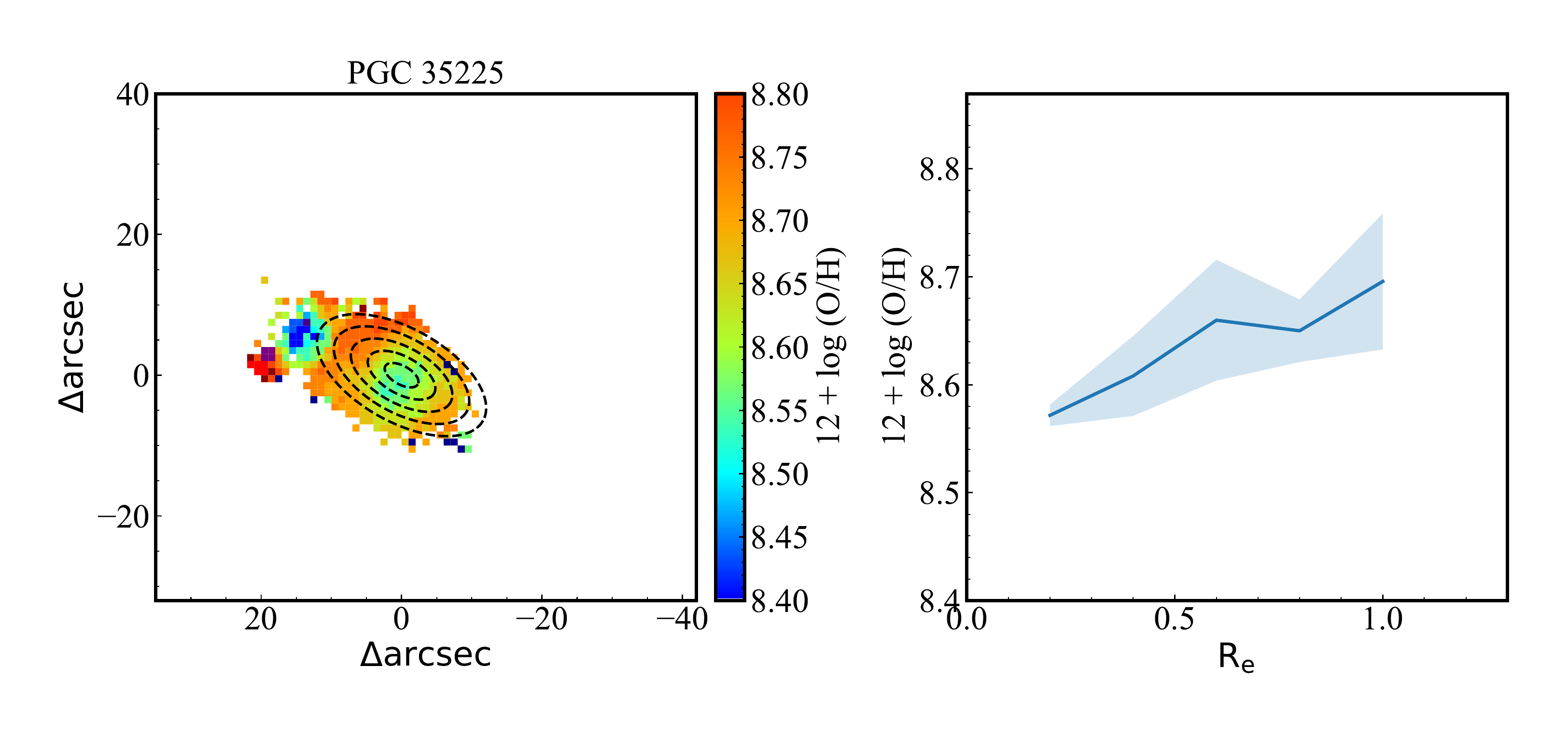}
\includegraphics[angle=0,width=0.42\textwidth]{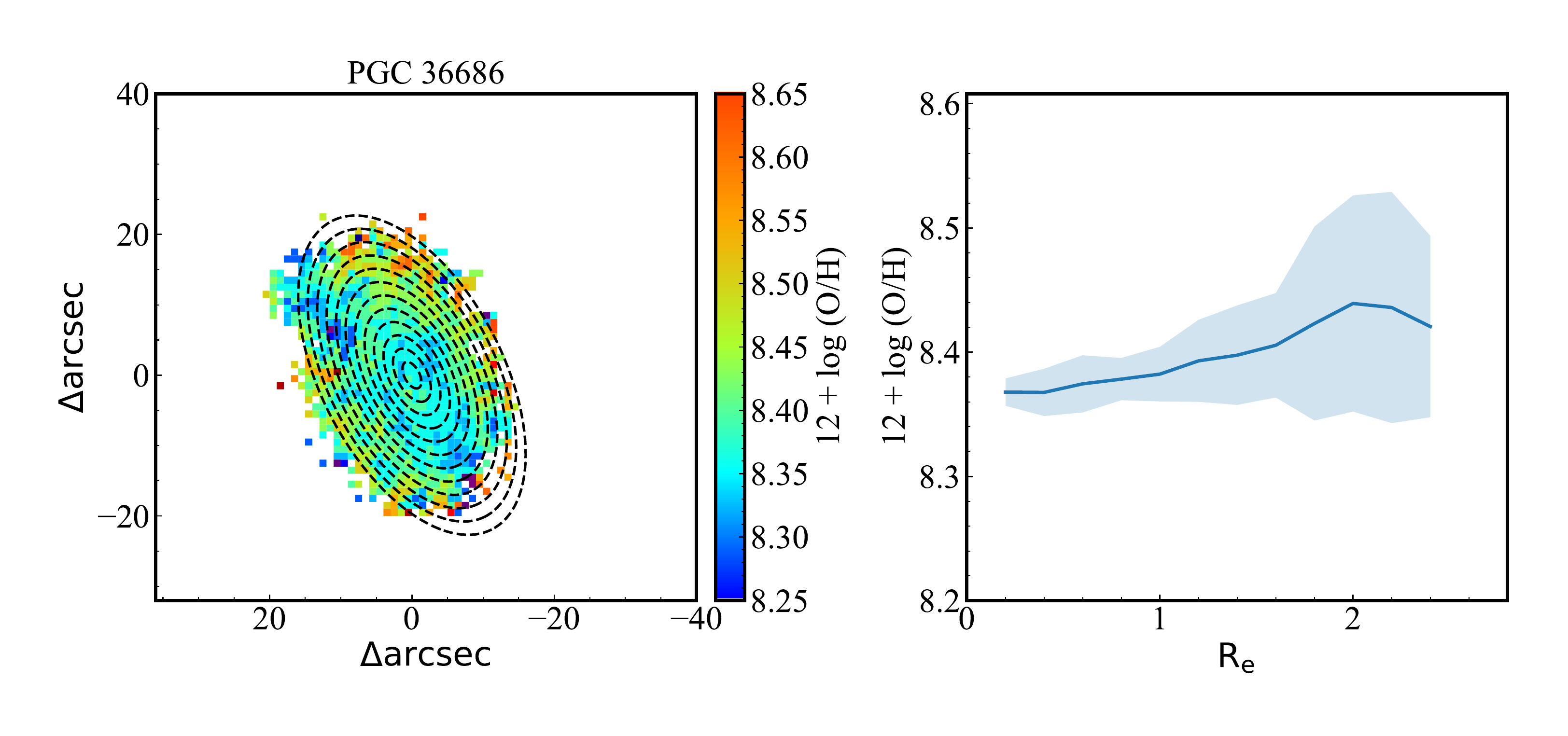}
\includegraphics[angle=0,width=0.42\textwidth]{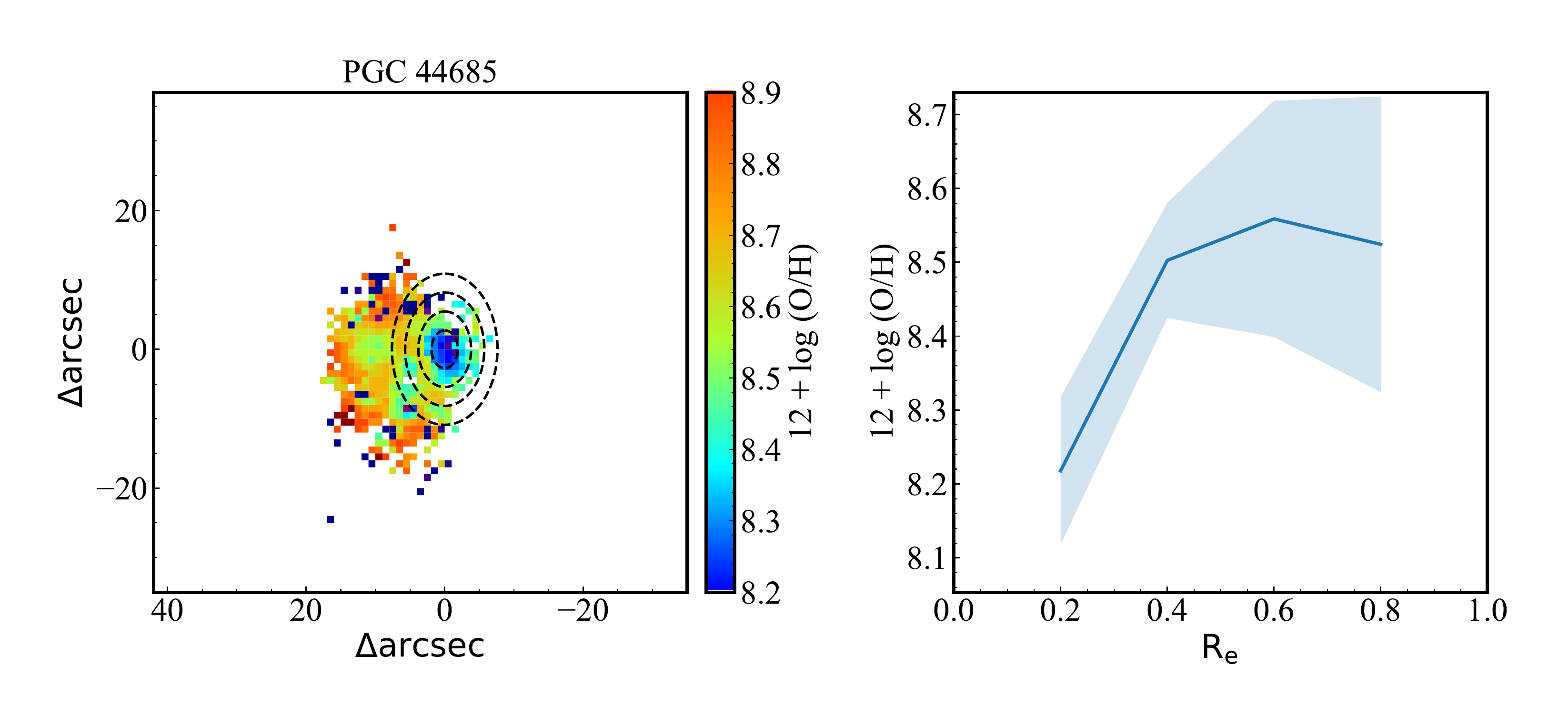}
\caption{Left in each panel: spatially resolved oxygen abundance for PGC 32543, PGC 35225, PGC 36686 and PGC 44685 for spaxels with S/N higher than 10 , respectively. Right in each panel:  radial distributions of oxygen abundance and error (the average oxygen abundance and the standard deviation in each circle.)}
\label{f11}
\end{figure*}

The metal content is a fundamental tracer of the star formation history. \cite{2020MNRAS.498.1939G} studied two star-forming dwarf galaxies in the Virgo cluster and found that these galaxies show positive gas metallicity gradients. They suggested that the dilution of gas-phase metallicity is associated with the recent merging event. In Figure \ref{f11}, we show the spatially resolved oxygen abundance and the radial distributions of the oxygen abundance for our sample. We find that the oxygen abundance in the star-forming regions are lower than that of the outside area. 
Although the oxygen abundance estimates have large errors, we also find from the radial distributions of the oxygen abundance that the oxygen abundance of these S0 galaxies increases with the increase in radius, showing a trend of positive metallicity gradients.
This result is consistent with \cite{2020MNRAS.498.1939G}.  Combined with the multi-nuclear structure and rich atomic gas, we suggest that the hierarchical accumulation paradigm might be an indication for the star formation in very low-mass S0 galaxies.

\section{Summary}
\label{sec5}

In this paper, we focus on the spatially resolved properties and cold gas content of four dwarf S0 galaxies with multiple cores based on the optical IFS observation from CAHA and the millimeter observation from IRAM-30 m. The main results are summarized as follows:

1. We find that all the sources in our sample (except PGC 35225, which have the highest ratio of molecular gas to stellar mass) have deviated from SFMS relation or are in the process of deviating from this relation. However, these S0 galaxies follow the K-S law within the star-forming region, which indicates that the star formation of dwarf S0 galaxies may be more extended.

2. In the velocity field distribution of star and ionized gas, we do not find regular rotating structures. Such a disturbed kinematics suggests that star formation in these S0 galaxies may be triggered by galaxy mergers or other external environmental effects. This picture is consistent with the multiple star-forming knots in our galaxies. We find that the star-forming knots have lower velocity dispersions of stars and ionized gas, which indicates that dwarf S0 galaxies still have significant rotation. The result is supported by the fact that most of these dwarf S0 galaxies are classified as fast rotators.

3. The physical properties of molecular gas of these dwarf S0 galaxies are very complicated. Their line ratios $I_{21}/I_{10}$ are all less than 1.0, which indicates that the distributions and the excitation temperatures of different CO transitions are very different.

4. We have detected molecular gas in all of these S0 galaxies, with molecular gas accounting for between 0.8\% and 1.5\% of the stellar mass. Although the content of molecular gas is very limited, we find that they store a lot of atomic gas, whose mass is almost comparable to stellar mass.
In addition, the relation between the gas-phase metallicities and SFR shows an inverse correlation, i.e., the richer the metal content, the lower the SFR. We suggest that
abundant atomic gas in these low-mass S0 galaxies may come from a series of merger and/or accretion events.
It was these rich atomic gas, combined with long dynamical time scale (fast rotator), that has led to the extended star formation history in these dwarf S0 galaxies of field environment. However, further observations are needed to confirm whether these atomic gas reservoirs belong to these galaxies.


\section*{Acknowledgements}
We are very grateful to the anonymous referee for critical comments and instructive suggestions, which significantly strengthened the analyses in this work.
This work is supported by the National Key Research and Development Program of China (No. 2017YFA0402703) and by the National Natural Science Foundation of China (No. 11733002). In addition, we acknowledge the supports of the staff from CAHA and IRAM-30 m. 
Rub\'en Garc\'ia-Benito acknowledges financial support from the Spanish Ministry of Economy and Competitiveness through grant 205 AYA2016-77846-P. Rub\'en Garc\'ia-Benito acknowledges support from the State Agency for 206 Research of the Spanish MCIU through the ``Center of Excellence Severo Ochoa'' award to 207 the Instituto de Astrof\'isica de Andaluc\'ia (SEV-2017-0709). We are grateful for the support of scientific research fund of Jiangsu Second Normal University (927801/032) and 
Yunnan Provincial Education Department (2021J0715).

\section*{Data Availability}
The data underlying this article cannot be shared publicly due to the privacy of individuals that participated in the study. The data will be shared on reasonable request to the other authors.

\end{document}